\documentclass[12pt,a4paper,fleqn]{article}

\usepackage{fullpage}

\usepackage{amsmath}
\usepackage{amssymb}
\usepackage{amscd}
\usepackage{plain}
\usepackage{graphicx}
\usepackage{epsfig}
\usepackage{fancyhdr}
\usepackage{t1enc,epsfig}
\usepackage[mathscr]{eucal}
\usepackage{epstopdf}
\usepackage[german,english]{babel}

\newtheorem{theorem}{Theorem}[section]

\newtheorem{corollary}[theorem]{Corollary}
\newtheorem{definition}[theorem]{Definition}
\newtheorem{example}[theorem]{Example}
\newtheorem{lemma}[theorem]{Lemma}
\newtheorem{proposition}[theorem]{Proposition}
\newtheorem{remark}[theorem]{Remark}
\newtheorem{conjecture}[theorem]{Conjecture}
\numberwithin{equation}{section}

\newenvironment{proof}{{\bf Proof. }}{\hfill$\rule{1ex}{1ex}$\par\medskip}

\begin{document}

\newcommand{\bt}{\begin{theorem}}
\newcommand{\et}{\end{theorem}}
\newcommand{\bd}{\begin{definition}}
\newcommand{\ed}{\end{definition}}
\newcommand{\bs}{\begin{proposition}}
\newcommand{\es}{\end{proposition}}
\newcommand{\bp}{\begin{proof}}
\newcommand{\ep}{\end{proof}}
\newcommand{\be}{\begin{equation}}
\newcommand{\ee}{\end{equation}}
\newcommand{\ul}{\underline}
\newcommand{\br}{\begin{remark}}
\newcommand{\er}{\end{remark}}
\newcommand{\bex}{\begin{example}}
\newcommand{\eex}{\end{example}}
\newcommand{\bc}{\begin{corollary}}
\newcommand{\ec}{\end{corollary}}
\newcommand{\bl}{\begin{lemma}}
\newcommand{\el}{\end{lemma}}
\newcommand{\bj}{\begin{conjecture}}
\newcommand{\ej}{\end{conjecture}}

\def\cmp{{\complement}}

\def\tA{{\tt A}}
\def\tB{{\tt B}}
\def\tC{{\tt C}}
\def\tD{{\tt D}}
\def\td{{\tt d}}
\def\tE{{\tt E}}
\def\tte{{\tt e}}
\def\tF{{\tt F}}
\def\tG{{\tt G}}
\def\tg{{\tt g}}
\def\ti{{\tt i}}
\def\tI{{\tt I}}
\def\tj{{\tt j}}
\def\tn{{\tt n}}
\def\tL{{\tt L}}
\def\tO{{\tt O}}
\def\tP{{\tt P}}
\def\tq{{\tt q}}
\def\ttr{{\tt r}}
\def\tP{{\tt P}}
\def\tR{{\tt R}}
\def\tS{{\tt S}}
\def\ttt{\tt t}
\def\tT{{\tt T}}
\def\ttg{{\tt g}}
\def\ttG{{\tt G}}
\def\bttg{\overline{\tg}}
\def\tU{{\tt U}} \def\tu{{\tt u}}
\def\tv{{\tt v}}
\def\tV{{\tt V}}
\def\tw{{\tt w}}
\def\tx{{\tt x}}
\def\ty{{\tt y}}
\def\tz{{\tt z}}

\def\bgam{{\mbox{\boldmath$\gamma$}}}
\def\uGam{\underline\Gamma}

\def\boeta{{\mbox{\boldmath$\eta$}}}
\def\oboeta{\overline\boeta}
\def\ups{\upsilon}

\def\Om{\Omega}
\def\om{\omega}
\def\oom{\overline\omega}
\def\bttg{\mbox{\boldmath${\tt g}$}}
\def\btau{\mbox{\boldmath${\tau}$}}
\def\bom{{\mbox{\boldmath$\omega$}}}
\def\obom{\overline\bom}
\def\0bom{{\bom}^0}
\def\0obom{{\obom}^0}
\def\nbom{{\bom}_n}
\def\0nbom{{\bom}_{n,0}}
\def\n*bom{{\bom}^*_{(n)}}
\def\wt{\widetilde}
\def\wtbom{\widetilde\bom}
\def\whbom{\widehat\bom}
\def\oom{\overline\om}
\def\wtom{\widetilde\om}
\def\bOm{\mbox{\boldmath${\Om}$}}
\def\obOm{\overline\bOm}
\def\whbOm{\widehat\bOm}
\def\wtbOm{\widetilde\bOm}

\def\Gam{\Gamma}
\def\Lam{\Lambda}
\def\lam{\lambda}

\def\Ups{\Upsilon}
\def\utheta{\underline\theta}
\def\ovr{\overline r}

\def\oG{\overline G}
\def\oL{\overline L}
\def\udelta{\underline\delta}
\def\uphi{\underline\Phi}
\def\upsi{\underline\Psi}

\def\bbC{\mathbb C}
\def\bbE{\mathbb E}
\def\bbP{\mathbb P}
\def\fB{\mathfrak B}
\def\fG{\mathfrak G}
\def\fW{\mathfrak W}
\def\bbQ{\mathbb Q}

\def\bi{\mathbf i}
\def\bj{\mathbf j}
\def\bn{\mathbf n}
\def\bbt{\mathbf t}
\def\bu{\mathbf u}
\def\bw{\mathbf w}
\def\bX{\mathbf X}
\def\ubX{\underline\bX}
\def\bx{\mathbf x}
\def\ubx{\underline\bx}
\def\bY{\mathbf Y}
\def\by{\mathbf y}
\def\ubY{\underline\bY}
\def\bZ{\mathbf Z}
\def\bz{\mathbf z}

\def\cl{\centerline}

\def\cA{\mathcal A}
\def\cB{\mathcal B}
\def\cC{\mathcal C}
\def\cD{\mathcal D}
\def\cE{\mathcal E}
\def\cF{\mathcal F}
\def\cH{\mathcal H}
\def\cK{\mathcal K}
\def\cL{\mathcal L}
\def\cN{\mathcal N}
\def\cS{\mathcal S}
\def\cT{\mathcal T}
\def\cV{\mathcal V}
\def\cW{\mathcal W}
\def\ocH{\overline\cH}
\def\ocW{\overline\cW}

\def\bbB{\mathbb B}
\def\bbK{\mathbb K}
\def\bbL{\mathbb L}

\def\bbR{\mathbb R}
\def\bbS{\mathbb S}
\def\bbT{\mathbb T}
\def\bbZ{\mathbb Z}
\def\ba{\mathbf a}
\def\bg{\mathbf g}
\def\bX{\mathbf X}
\def\bx{\mathbf x}
\def\wtbx{\widetilde\bx}
\def\ui{{\underline i}}

\def\oA{{\overline A}}
\def\uA{{\underline A}}
\def\ua{{\underline a}}
\def\uua{{\underline{a_{}}}}
\def\oa{{\overline a}}
\def\uk{{\underline k}}
\def\ux{{\underline x}}
\def\wtux{\widetilde\ux}
\def\uX{{\underline X}}
\def\by{\mathbf y}
\def\uy{\underline y}
\def\bY{\mathbf Y}
\def\uY{\underline Y}

\def\uj{{\underline j}}
\def\unn{\underline n}
\def\unp{\underline p}
\def\ovp{\overline p}
\def\bx{\mathbf x}
\def\ox{\overline x}
\def\obx{\overline\bx}
\def\uz{\underline z}
\def\bz{\mathbf z}
\def\uv{\underline v}
\def\dist{\textrm{dist}}
\def\diy{\displaystyle}
\def\ov{\overline}
\def\u0{{\underline 0}}

\def\oomega{\overline\omega}
\def\oUpsilon{\overline\Upsilon}
\def\wtomega{\widetilde\omega}
\def\wtz{\widetilde z}
\def\wtheta{\widetilde\theta}
\def\wtalpha{\widetilde\alpha}
\def\wh{\widehat}
\def\oV{\overline {\mathcal V}}

\def\bI{\mathbf I}
\def\bN{\mathbf N}
\def\bbN{\mathbb N}
\def\bP{\mathbf P}
\def\bV{\mathbf V}
\def\oW{\overline W}
\def\ofW{\overline\fW}
\def\LT{{\mathbb{LT}}}
\def\mucr{{\mu_{cr}}}

\def\rA{{\rm A}}  \def\rB{{\rm B}}
\def\orB{\overline\rB} \def\urB{\underline\rB}
\def\rc{{\rm c}}  \def\rC{{\rm C}}
\def\rd{{\rm d}}  \def\rD{{\rm D}}
\def\re{{\rm e}}  \def\rE{{\rm E}}
\def\rF{{\rm F}}
\def\rI{{\rm I}}

\def\rn{{\rm n}}

\def\rO{{\rm O}}
\def\rP{{\rm P}}
\def\rQ{{\rm Q}} \def\urQ{\underline\rQ}
\def\rr{{\rm r}}
\def\rR{{\rm R}} \def\urR{\underline\rR}

\def\rs{{\rm s}}
\def\rS{{\rm S}}
\def\rT{{\rm T}}
\def\rV{{\rm V}}

\def\rw{{\rm w}}

\def\rx{{\rm x}}
\def\ry{{\rm y}}
\def\rtr{\rm{tr}}

\def\oa{\overline a}
\def\ua{\underline a}

\def\uk{\underline k}
\def\un{\underline n}
\def\ux{\underline x}
\def\uy{\underline y}
\def\wtux{\widetilde\ux}
\def\uX{\underline X}

\def\oJ{\overline J}
\def\oP{\overline P}
\def\utC{{\underline\tC}}
\def\utD{{\underline\tD}}
\def\utE{{\underline\tE}}
\def\urB{{\underline\rB}}
\def\urC{{\underline\rC}}
\def\urD{{\underline\rD}}
\def\urE{{\underline\rE}}

\def\oXi{{\overline\Xi}}

\def\vng{{\varnothing}}
\def\cl{\centerline}

\def\BX{{\mathbf X}} \def\bx{{\mathbf x}}
\def\BY{{\mathbf Y}} \def\by{{\mathbf y}}

\def\bbZ{{\mathbb Z}} \def\bbP{{\mathbb P}}
\def\bz{\mathbf z}

\def\cA{{\mathcal A}}
\def\cB{{\mathcal B}} \def\cX{{\mathcal X}}
\def\bcA{{\mbox{\boldmath{$\cA$}}}}
\def\bcX{{\mbox{\boldmath{$\cX$}}}}

\def\fB{\mathfrak B}\def\fM{\mathfrak M} \def\fX{\mathfrak X}
\def\cT{\mathcal T}
\def\bu{\mathbf u}
\def\bv{\mathbf v}\def\bx{\mathbf x}\def\by{\mathbf y}
\def\om{\omega} \def\Om{\Omega}
\def\bbP{\mathbb P} \def\hw{{h^{\rm w}}} \def\hwphi{{h^{\rm w}_\phi}}
\def\beq{\begin{eqnarray}} \def\eeq{\end{eqnarray}}
\def\beqq{\begin{eqnarray*}} \def\eeqq{\end{eqnarray*}}

\def\rb{{\rm b}}
\def\rd{{\rm d}} \def\rmv{{\rm v}} \def\rV{{\rm V}}
\def\Dwphi{{D^{\rm w}_\phi}}
\def\BX{\mathbf{X}}
\def\hwphiii{{h^{\rm w}_{\phi_1\otimes\phi_2\otimes\dots\otimes \phi_n}}}
\def\hwphii{{h^{\rm w}_{\phi_1\otimes\phi_2}}}
\def\mwe{{D^{\rm w}_\phi}}
\def\DwPhi{{D^{\rm w}_\Phi}} \def\iw{i^{\rm w}_{\phi}}
\def\bE{\mathbb{E}} \def\mbE{\mathbf{E}}
\def\mbp{{\mathbf p}}
\def\1{{\mathbf 1}} \def\fB{{\mathfrak B}}  \def\fM{{\mathfrak M}}
\def\bbE{{\mathbb E}}

\def\tha{{\theta}} \def\uBX{{\underline\BX}}
\def\gam{\gamma} \def\kap{\kappa} \def\lam{\lambda}
\def\ups{{\upsilon}}

\def\vphi{\varphi} \def\vpi{\varpi}
\def\veps{\varepsilon} \def\vrho{\varrho}
\def\vtheta{\vartheta} \def\vpi{\varpi}
\def\ovphi{\overline\vphi}

\def\ot{\leftarrow}

\def\blam{{\mbox{\boldmath${\lambda}$}}} \def\bphi{{\mbox{\boldmath${\phi}$}}}
\def\bpsi{{\mbox{\boldmath${\psi}$}}} \def\bta{{\mbox{\boldmath${\eta}$}}}
\def\bzeta{{\mbox{\boldmath${\zeta}$}}} \def\btau{{\mbox{\boldmath${\tau}$}}}
\def\bups{{\mbox{\boldmath${\ups}$}}}
\def\bUps{{\mbox{\boldmath${\Ups}$}}}
\def\bu{\mathbf u}
\def\bU{\mathbf U}
\def\bT{\mathbf T}
\def\bpi{{\mbox{\boldmath${\pi}$}}}

\def\lam{{\lambda}} \def\eps{{\epsilon}}
\def\Lam{{\Lambda}} \def\BK{{\mathbf K}} \def\BP{{\mathbf P}}
\def\bOne{\mathbf 1} \def\uOne{\underline 1} \def\ulam{{\underline\lam}}
\def\bbT{\mathbb T}

\def\urM{{\underline{\rm M}}}
\def\uPi{{\underline\Pi}}
\def\rC{{\rm C}} \def\rL{{\rm L}}  \def\rM{{\rm M}}
\def\rP{{\rm P}} \def\rR{{\rm R}} \def\cA{\mathcal A}
 \def\cB{\mathcal B} \def\cF{\mathcal F} \def\cG{\mathcal G}
\def\cM{\mathcal M} \def\cP{\mathcal P} \def\cV{\mathcal V}
\def\cX{\mathcal X} \def\cY{\mathcal Y}

\def\rt{{\rm t}} \def\ru{{\rm u}}
\def\rv{{\rm v}} \def\rw{{\rm w}}

\def\bbR{{\mathbb R}}
\def\bbA{{\mathbb A}} \def\bbB{{\mathbb B}}
\def\bbC{{\mathbb C}}  \def\bbE{{\mathbb E}}
\def\bbD{{\mathbb D}}\def\bbG{{\mathbb G}}
\def\bbM{{\mathbb M}} \def\bbP{\mathbb P}
 \def\bbS{{\mathbb S}}
  \def\bbZ{{\mathbb Z}}

\def\ree{{\rm e}}

\def\tA{{\tt A}} \def\tB{{\tt B}} \def\tC{{\tt C}}
\def\tf{{\tt f}}  \def\tg{{\tt g}} \def\thh{{\tt h}}
\def\tI{{\tt I}}
\def\tJ{{\tt J}} \def\tK{{\tt K}}
\def\tL{{\tt L}} \def\tP{{\tt P}} \def\tp{{\tt p}}
\def\tq{{\tt q}} \def\tQ{{\tt Q}}
\def\tR{{\tt R}} \def\tS{{\tt S}}
\def\tW{{\tt W}}
\def\ty{{\tt y}} \def\tz{{\tt z}}
\def\t0{{\tt 0}} \def\tp{{\tt 1}}

\def\fy{{\mathfrak y}} \def\fz{{\mathfrak z}}

\def\iGam{{\mathit{\Gam}}}
\def\igam{{\mathit{\gamma}}}
 \def\iUps{{\mathit{\Ups}}} \def\ups{{\upsilon}}
\def\iups{{\mathit{\ups}}}
\def\itau{{\mathit{\tau}}}

\def\be{\begin{equation}}
\def\ee{\end{equation}}

\def\rL{\rm L} \def\rM{\rm M}
\def\rN{\rm N}

\def\beal{\begin{array}{l}}
\def\beac{\begin{array}{c}}
\def\bear{\begin{array}{r}}
\def\beacl{\begin{array}{cl}}
\def\beacr{\begin{array}{cr}}
\def\ena{\end{array}}

\def\diy{\displaystyle}

\def\sA{\mathscr A} \def\sB{\mathscr B} \def\sC{\mathscr C}
\def\sF{\mathcal F} \def\sG{\mathcal G} \def\sI{\mathcal I}
\def\sL{\mathcal L}\def\sM{\mathscr M} \def\sO{\mathcal O}
\def\sP{\mathcal P} \def\sR{\mathcal R} \def\sS{\mathcal S}
\def\oB{{\overline B}}

\def\ri{{\rm i}} \def\BC{{\mathbf C}} \def\u1{{\underline 1}} \def\uv{\underline v}\def\Bf0{\mathbf 0}
\def\opsi{{\overline\psi}} \def\ovp{{\overline p}} \def\tp{{\tt p}}

\title{\bf Weighted information and entropy rates}

\author{\bf Y. Suhov$^1$, I. Stuhl$^2$}

\date{}
\footnotetext{ \vskip 3 truemm

\noindent $^1$ Mathematics Dept., Penn State University, University Park, State College,
PA 16802, USA; DPMMS, University of Cambridge, UK;\\
E-mail: yms@statslab.cam.ac.uk

\noindent $^2$ Mathematics Dept.,
University of Denver, Denver, CO 80208 USA;
 Appl. Math and Prob. Theory Dept., University of Debrecen, Debrecen, 4028, HUN;\\
E-mail: izabella.stuhl@du.edu}

\maketitle

\begin{abstract}
The weighted entropy $H^\rw_\phi (X)=H^\rw_\phi (f)$ of a random variable
$X$ with values $x$ and a probability-mass/density function $f$ is defined as
the mean value $\bbE I^\rw_\phi(X)$ of the weighted information
$I^\rw_\phi (x)=-\phi (x)\log\,f(x)$. Here $x\mapsto\phi (x)\in\bbR$ is a given weight function
(WF) indicating a 'value' of outcome $x$. For an $n$-component random vector
$\BX_0^{n-1}=(X_0,\ldots ,X_{n-1})$ produced by a random process
$\BX=(X_i,i\in\bbZ)$, the weighted information $I^\rw_{\phi_n}(\bx_0^{n-1})$ and weighted
entropy $H^\rw_{\phi_n}(\BX_0^{n-1})$
are defined similarly, with an WF $\phi_n(\bx_0^{n-1})$. Two types
of WFs $\phi_n$ are considered, based on additive and a multiplicative
forms ($\phi_n(\bx_0^{n-1})=\sum\limits_{i=0}^{n-1}\vphi (x_i)$ and
$\phi_n(\bx_0^{n-1})=\prod\limits_{i=0}^{n-1}\vphi (x_i)$, respectively). The focus is upon
{\it rates} of the weighted entropy and information, regarded as parameters related to
$\BX$. We show that, in the context of ergodicity, a natural scale for an
asymptotically additive/multiplicative WF is $\diy\frac{1}{n^2}H^\rw_{\phi_n}(\BX_0^{n-1})$
and $\diy\frac{1}{n}\log\;H^\rw_{\phi_n}(\BX_0^{n-1})$, respectively. This
gives rise to {\it primary rates}. The next-order terms can also be identified, leading to
{\it secondary rates}. We also consider emerging generalisations of the Shannon--McMillan--Breiman
theorem.
\end{abstract}

\section{Introduction}

The purpose of this paper is to introduce and analyze weighted entropy rates for some
basic random processes. In the case of a standard entropy, the entropy rate is a
fundamental parameter
leading to profound results and fruitful theories with far-reaching
consequences, cf. \cite{CT}. The case of weighted entropies is much less developed,
and this paper attempts to cover a number of aspects of this notion.
In this work we treat two types of weight functions: additive and multiplicative.
Conceptually, the present paper continues Refs \cite{SSYK, SSY} and is connected with
\cite{SS}.
% The rates of weighted information/entropy are defined in Eqns \eqref{eq:rateAdd},
% \eqref{rhom} at the end of Section 2.

We work with a complete probability space $(\Om ,\fB,\bbP)$ %(see, e.g., \cite{I})
and consider
random variables (RVs) as (measurable) functions $\Om\to\cX$ taking values in
a measurable space $(\cX,\fM )$ equipped with a countably additive reference measure $\nu$.
Probability mass functions (PMFs) or probability density functions (PDFs) are defined
relative to $\nu$. (The difference between PMFs (discrete parts of probability measures)
and PDFs (continuous parts) is insignificant for most of the work;
this will be reflected in a common acronym PM/DF.)  In the case of  an RV collection $\{X_i\}$, the
space of values $\cX_i$ and the reference measure $\nu_i$ may vary with $i$. (Some of the
$X_i$ may be random vectors.)

Given a (measurable) function $x\in\cX\mapsto\phi (x )\in\bbR$, and an RV
$X:\;\Om\to\cX$, with a PM/DF $f$, the weighted information (WI) $I^\rw_{\phi} (x)$
with  weight function (WF) $\phi$ contained in an outcome $x\in\cX$ is given by
\be\label{eq:Iwphi} I^\rw_{\phi}(x)=-\phi (x)\log f (x).\ee
The symbol $I^\rw_{\phi}(X)$ is used for the random WI, under PM/DF $f$.
Next, one defines
the weighted entropy (WE ) $h^\rw_\phi (f)$ of $f$ (or $X$) as
\be\label{eq:1.1}h^\rw_\phi (f) =-\int_\cX\phi (x )f(x )\log\,f(x)\nu (\rd x )=
\bbE\,I^\rw_\phi (X)\ee
whenever the integral $\int_\cX |\phi (x )|\,f(x )|\log\,f(x)|\nu (\rd x )<\infty$. (A
common agreement $0=0\cdot\log\,0=0\cdot\log\,\infty$ is in place throughout the paper.)
Here and below we denote by $\bbE$ the expectation relative to $\bbP$ (or
an induced probability measure emerging in a given context).
For $\phi (x)\geq 0$, the WE  in a discrete case (when $\cX$ is a finite or a countable set)
is non-negative. For $\phi (x)=1$, we obtain the
standard information $I(x)=-\log\,f(x)$ (SI) and standard entropy $h(f)=\bbE\,I(X)$ (SE).

Let $\BX_0^{n-1}=(X_0,X_1,\dots,X_{n-1})$ be a random vector (string), with components
$X_i:\;\Om\to\cX_i$, $0\leq i\leq n-1$. Let $f_n(\bx_0^{n-1})$ be the joint PM/DF
relative to measure $\nu_0^{n-1}(\rd\bx_0^{n-1})=\prod\limits_{i=0}^{n-1}\nu_i(\rd x_i)$ where
$\bx_0^{n-1}=(x_0,\ldots ,x_{n-1})\in\operatornamewithlimits{\times}\limits_{i=0}^{n-1}\cX_i:=\cX_0^{n-1}$.
Given a function $\bx_0^{n-1}\in\cX_0^{n-1}\mapsto\phi_n(\bx_0^{n-1} )\in\bbR$,
the joint WE $h^\rw_{\phi_n}(f_n)$ of $X_0,\ldots ,X_{n-1}$ with WF $\phi_n$ is given by
\be\label{eq:1.3a}h^\rw_{\phi_n}(f_n)=-\int_{\cX_0^{n-1}}\phi_n(\bx_0^{n-1} )f_n(\bx_0^{n-1})\log\,
f_n(\bx_0^{n-1})\nu_0^{n-1}(\rd\bx_0^{n-1} )=\bbE\,I^\rw_{\phi_n}(\BX_0^{n-1}),\ee
where $I^\rw_{\phi_n}(\bx_0^{n-1})$ represents the WI in the joint outcome $\bx_0^{n-1}$:
\be\label{eq:1.3b}I^\rw_{\phi_n}(\bx_0^{n-1})=-\phi_n(\bx_0^{n-1})\log\,f_n(\bx_0^{n-1}).\ee

We focus upon two kinds of weight functions $\phi_n(\BX_0^{n-1})$: additive
and multiplicative, and their asymptotical modifications.
Both relate to the situation where $\BX_0^{n-1}=(X_0,\ldots ,X_{n-1})$ and each component
$X_j$ takes values in the same space: $\cX_j=\cX_1=\cX$. In the simplest form, additivity and
multiplicativity mean representations
\be\label{eq:1.4}\phi_n(\bx_0^{n-1})=\sum\limits_{0\leq j < n}\vphi (x_j)\;\hbox{ and }\;
\phi_n(\bx_0^{n-1})=\prod\limits_{0\leq j < n}\vphi (x_j),\ee
where $x\in\cX\mapsto\vphi (x)$ is a given functions (one-digit WFs). In
the additive case we can allow $\vphi$ to be of
both signs whereas in the multiplicative case we suppose $\vphi \geq 0$.
% A convenient generalization of
% \eqref{eq:1.4} is a concept of  asymptotic additivity/multiplicativity in various forms;
% this is discussed in the next section.

Additive weight functions may emerge in relatively stable situations where each observed digit $X_j$
brings reward or loss $\vphi (X_j)$ (bearing opposite signs); the summatory value $\phi_n(\BX_0^{n-1})$
is treated as a cumulative gain or deficit after $n$ trials. Multiplicative weight functions reflect a
more turbulent scenario where
the value $\phi_n(\BX_0^{n-1})$ increases/decreases by a factor $\vphi (X_n)$ when outcome
$X_n$ is observed. Cf. \cite{SSK}. As before, for $\phi_n(\bx_0^{n-1})\equiv 1$ we obtain the
SE $h(f_n)$ and SI $I(\bx_0^{n-1})$.

Our goal is to introduce concepts of {\it rates} for $h^\rw_{\phi_n}(f_n)$ and $I^\rw_{\phi_n}(\bx_0^{n-1})$
characterising the order of growth/decay as $n\to\infty$. To this end we consider
a (discrete-time) random process $\BX=(X_i,i\in\bbZ)$ or $\BX=(X_i,i\in\bbZ_+)$,
with a probability distribution $\bbP$;
vector $\BX_0^{n-1}$ will represent an initial string  generated by the process.  In the case of the
SE and SI, the rates are defined as $\lim\limits_{n\to\infty}\diy\frac{1}{n}h(f_n)$ and
$\lim\limits_{n\to\infty}\diy\frac{1}{n}I(\bx_0^{n-1})$, and for an ergodic process they coincide
almost everywhere relative to the distribution $\bbP$. See \cite{AC}, \cite{Ba}, \cite{CT}. For
the WE and WI we find it natural to introduce {\it primary} and {\it secondary} rates. The former
emerges as a limit of \ $\diy\frac{1}{n^2}H^\rw_{\phi_n}(\BX_0^{n-1})$ for asymptotically additive
WFs and of \ $\diy\frac{1}{n}\log\;H^\rw_{\phi_n}(\BX_0^{n-1})$ for asymptotically multiplicative WFs.
The secondary rate, roughly, provides a `correction term', although in a number of situations
(when the primary rate vanishes) the secondary rate should bear a deeper significance.
We also consider generalisations of the Shannon--McMillan--Breiman (SMB) theorem for
asymptotically additive WFs.

The paper is organised as follows. In Section 2 we put forward the concepts of asymptotically
additive and multiplicative WFs. In Section 3, the primary and secondary rates for additive case
are discussed. Section 3 ...

\section{Asymptotic additivity and multiplicativity}

Here we introduce classes of asymptotically additive and multiplicative WFs
for which we develop results on rates in the subsequent sections. The object of study is
a discrete-time random process $\BX_0^\infty =(X_n:\;n\in\bbZ_+)$ or $\BX=(X_n:\;n\in\bbZ)$.
We begin with a simple example
where $\BX_0^\infty$ is an IID
(Bernoulli) process with values in $\cX$: here, for $\bx_0^{n-1}=(x_0,\ldots ,x_{n-1})\in\cX^n$, the
joint PM/DF for string $\BX_0^{n-1}=(X_0,\ldots ,X_{n-1})$ is $f_n(\bx_0^{n-1})=
\prod\limits_{i=0}^{n-1}p(x_i)$ where $p(x)=p_0(x)$ is the one-time marginal PM/DF, $x\in\cX$.
We start with a straightforward remark:
\begin{description}
  \item[{\rm{(a)}}] For a sequence of IID random variables $\BX_0^\infty$ and an additive WF \
$\phi_n (\BX_0^{n-1})$ \\ $=\sum\limits_{0\leq j<n}\vphi (X_j)$, the WI has a representation:
\be\label{eq:SL1}I^\rw_{\phi_n}(\BX_0^{n-1})=-\phi (\BX_0^{n-1})\log\,f_n(\BX_0^{n-1})
=-\sum\limits_{j=0}^{n-1}\vphi (X_j)\sum\limits_{l=0}^{n-1}\log\,p(X_l).\ee
Next, with $H(p)=-\bbE\big[\log p(X)\big]$ and
$H^\rw_\vphi (p)=-\bbE\big[\vphi (X)\log p(X)\big]$ (the one-digit SE and WE, respectively):
 \be\label{eq:L1}H^{\rm w}_{\phi_n^{\,}}(f_n)=n(n-1)H(p) \bbE\big[\vphi (X)\big]+nH^\rw_\vphi (p)
:=n(n-1)\rA_0+n\rA_1.\ee
  \item[{\rm{(b)}}] For a sequence of IID random variables $\BX_0^\infty$ and a multiplicative WF
 $\phi_n (\BX_0^{n-1})$\\ $=\prod\limits_{0\leq j< n}\vphi (X_j)$:
\be\label{eq:SM1}I^\rw_{\phi_n}(\BX_0^{n-1})=-\phi_n(\BX_0^{n-1})\log\,f_n(\BX_0^{n-1})
=-\prod\limits_{j=0}^{n-1}\vphi (X_j)\sum\limits_{l=0}^{n-1}\log\,p(X_l). \ee
Next,
 \be\label{eq:M1} H^{\rm w}_{\phi_n^{\,}}(f_n)=nH^\rw_\vphi (p)\big[\bbE\vphi (X)\big]^{n-1}
:= \rB_0^{n-1}\times n\rB_1.\ee
\end{description}

Values $\rA_0$ and $\rB_0$ are referred to as primary rates and $\rA_1$ and $\rB_1$
as secondary rates. \vskip 3 truemm

Eqns \eqref{eq:SL1}--\eqref{eq:M1} %\eqref{eq:L1}, \eqref{eq:SM1} and
provide intuition for formulas of convergence \eqref{eq:rateAdd}--\eqref{rhom} which yield
versions of the SMB theorem for the WI and WE in a general case with asymptotically additive WFs.
(A number of
subsequent results will be established or illustrated under
specific restrictions, viz., Markovian or Gaussian assumptions.)  We consider
$\bcX=\cX^{\bbZ}$ (the space of trajectories
over $\bbZ$) and $\bcX_+=\cX^{\bbZ_+}$ (the set of trajectories over $\bbZ_+$),
equipped with the corresponding sigma-algebras.
% In Eqns \eqref{eq:rateAdd}--\eqref{rhom} deal with a general random process (RP)
% $\BX_0^\infty =(X_n:\,n\geq 0)$ or $\BX=(X_n:\;n\in\bbZ)$, specifying, when necessary, its
% particular structure.
As was said, symbol $\bbP$ is used for a
probability measure on $\bcX_+$  or $\bcX$ generated by process $\BX_0^\infty$ or
$\BX$. (In the case of $\BX$, symbol $\bbP$ will be related to a stationary process,
while for $\BX_0^\infty$ some alternative possibilities can be considered as well,
involving initial conditions.)
Symbol $\bbE$ refers to the expectation relative to $\bbP$. Next, $\rL_2$
stands for the Hilbert space $\rL_2(\bcX_+,\bbP)$ or $\rL_2(\bcX,\bbP)$
and $\rL_1$ for the space $\rL_1(\bcX_+,\bbP)$ or $\rL_1(\bcX,\bbP)$. The joint PM/DF for string
$\BX_0^{n-1}$ is again denoted by $f_n$: $f_n(\bx_0^{n-1})=\diy\frac{\bbP(\BX_0^{n-1}
\in\rd\bx_0^{n-1})}{\nu^n(\rd\bx_0^{n-1})}$.
The focus will be upon rates of the WI $I^\rw_{\phi_n}(\BX_0^{n-1})$ and WE
$H^\rw_{\phi_n}(f_n)$; see \eqref{eq:1.3b} and \eqref{eq:1.3a}.

One of aspects of this work is to outline
general classes of  WFs $\phi_n$ and RPs $\BX$, replacing the exact formulas in
\eqref{eq:L1} and \eqref{eq:M1} by
suitable asymptotic representations (with emerging asymptotic counterparts of parameters
$\rA_0$, $\rA_1$ and $\rB_0$, $\rB_1$). In our opinion, a natural class of RPs here are ergodic
processes; a part of the assertions in this paper are established in this class. The basis for
such a view is that for an ergodic RP $\BX=(X_n,\,n\in\bbZ)$ the limit
\be\label{eq:SMBGen}\lim\limits_{n\to\infty}\diy\frac{-1}{n}\log\,f_n(\BX_0^{n-1}) =h\ee
exists $\bbP$-a.s. according to results by Barron (1985) \cite{Ba} and Algoet--Cover (1988) \cite{AC}.
Cf., e.g., \cite{AC}, Theorem 2, and the biblio therein. The limiting value $h$ is identified
as the SE rate of RP $\BX$ (the SMB theorem).
%Consequently,
%we can use the above conditions \eqref{eq:genAdd} in an attempt to fit results and
%proofs from \cite{AC}. % For definiteness, we choose the first option, assuming \eqref{eq:conv1}.
However, a number of properties in the present paper
are proven under Markovian assumptions, due to technical complications. In some situations
(for Gaussian processes) we are able to analyse the situation without referring directly to
ergodicity (or stationarity).

Another aspect is related to suitable assumptions upon WFs. One assumption is that
\be\label{eq:genAdd}\lim\limits_{n\to\infty}\frac{1}{n}\phi_n (\BX_0^{n-1})=\alpha,\;\;\bbP\hbox{-a.s.}\;
\hbox{and/or in \ $\rL_2$}\;\;\hbox{(asymptotic additivity);}\ee
together with \eqref{eq:SMBGen} it leads to identification of the primary rate $\rA_0$:
\beq\label{eq:Aoiden} \rA_0=\alpha h.\eeq
The impact of process $\BX$ in assumption \eqref{eq:genAdd} is reduced to the form of convergence
($\bbP$-a.s. or $\rL_2(\bcX,\bbP)$). A stronger tie between $\phi_n$ and $\BX$ is introduced in
an asymptotic relation \eqref{eq:genAdd2} arising from \eqref{eq:L1}:
\be\label{eq:genAdd2}\beal\lim\limits_{n\to\infty}{\diy\frac{1}{n}}
H^\rw_{\phi_n}(f_n)=\rA_1.\ena\ee
% (Of course, \eqref{eq:genAdd2} is the same as $I^\rw_{\phi_n}(\BX_0^{n-1})/n\to\rA_1$
% removing any element of surprise. However, it remains to be verified for natural classes of
% RPs $\BX$ and WFs $\phi_n$.)
An instructive property implying \eqref{eq:genAdd2} is that
$\forall$ $j\in\bbZ$,
\be\label{eq:genAdd3}
\beal\lim\limits_{n\to\infty}\bbE\big[\phi_n (\BX_0^{n-1})\log\,p^{(j)}(X_j|\BX_0^{j-1})\big]=\rA_1
%\qquad\qquad\qquad
\quad\hbox{(strong asymptotic additivity).}\ena\ee
This yields an identification of the secondary rate $\rA_1$. Here and below,
$p^{(j)}(y|\bx_0^{j-1})$ represents the conditional PM/DF of having $X_j=y$ given that
string $\BX_0^{j-1}$ coincides with $\bx_0^{j-1}$; see Eqn \eqref{eq:3.2} below.
Assumptions \eqref{eq:genAdd} and \eqref{eq:genAdd3} are relevant in Section 3, Theorem \ref{Thm1}.

An informal meaning of \eqref{eq:genAdd} is that there is an approximation
\be\label{CondB}\beal{\diy\frac{\phi_n(\bx_0^{n-1})-\phi^*_n(\bx)}{n}}\to 0\;\hbox{ where }\;
%\qquad\qquad\qquad
\;\phi^*_n(\bx)=\sum\limits_{j=0}^{n-1}\vphi^*(S^j\bx),\quad\ena\ee
for some measurable function $\bx\in{\mbox{\boldmath{$\cX$}}}\mapsto\vphi^*(\bx)\in\bbR$ from
$\rL_1$, with $\alpha =\bbE\vphi^*(\BX)$. Here and below,
$S$ stands for the shift in $\bcX$: $(S^j\bx)_l=x_{l-j}$ for $\bx =(x_l)\in\bcX$.
% In analogy with Eqn \eqref{eq:L1}, parameter
% $\rA_0$ will be be identified as the product $h\alpha $ where $h$ is the entropy rate of process $\BX$.
% A simplifying assumption will be that $\phi_n(\bx_0^{n-1})=\sum\limits_{j=0}^{n-1}\vphi (x_j)$,
% with $\vphi^*(\bx)=\vphi (x_0)$; cf. \eqref{eq:1.4}.
From this point of view, condition \eqref{eq:genAdd2} is instructive when $\rA_0=0$ (i.e.,
$h$ or $\alpha$ vanishes).
% Similarly, \eqref{eq:genAdd3} means that the WI $I^\rw_{\phi_n}(\BX_0^{n-1})$
% behaves like the sum $\sum\limits_{j=0}^{n-1}\vtheta^*(S^j\BX)$ allowing us to use ergodicity
% and the law of large numbers:
% \be\label{CondB1}\beal\diy\frac{I^\rw_{\phi_n}(\bx_0^{n-1})-\theta^*_n(\bx)}{n}\to 0\;\hbox{ where }\;
% \qquad\qquad\qquad
% \;\theta^*_n(\bx)=\sum\limits_{j=0}^{n-1}\vtheta^*(S^j\bx).\quad\ena\ee

Let us now pass to multiplicative WFs. An assumption used in Section 4, Theorem 5, claims that
\be\label{eq:genMul}\beal\lim\limits_{n\to\infty}{\diy\frac{1}{n}}\log\phi_n (\BX_0^{n-1})
=\log\,\beta,\;\;\hbox{or, equivalently,}\\
\qquad \lim\limits_{n\to\infty}\big[\phi_n (\BX_0^{n-1})\big]^{1/n}=\beta,
\;\;\bbP\hbox{-a.s.}\;\;\hbox{(asymptotic multiplicativity).}\ena\ee
Similarly to \eqref{CondB}, Eqn \eqref{eq:genMul}
means, essentially, that
\be\label{CondC}\beal{\diy\left[\frac{\phi_n(\BX_0^{n-1})}{\phi^*_n(\BX)}\right]^{1/n}}\to 1\;\hbox{ where }\;
\;\phi^*_n(\bx)=\prod\limits_{0\leq j< n}\vphi^*(S^j\bx),\ena\ee
for some measurable function $\bx\in{\mbox{\boldmath{$\cX$}}}\mapsto\vphi^*(\bx)>0$, with
$(\log\,\vphi^*)\in\rL_1$ and $\bbE\log\,\vphi^*(\BX )=\beta$.  %Condition \eqref{CondC} is equivalent to
%\be\label{CondD}\left[\frac{\phi_n(\BX_0^{n-1})}{\phi^*_n(\BX_0^{n-1})}\right]^{1/n}\to 1,\;\;
%\bbP\hbox{-a.s.,\;\; as }\;n\to\infty ,\;\;\hbox{where $\phi^*_n(\bx)=\prod\limits_{0\leq j< n}\vphi^*(S^j\bx)$.}\ee
A stronger form of such a condition is an exact equality: $\phi_n(\bx_0^{n-1})=
\prod\limits_{0\leq j\l<n}\vphi (x_j)$; cf. \eqref{eq:1.4}.

For a future use, we suggest an integral form of condition \eqref{CondC}: as $n\to\infty$,
\be\label{eq:CondI}\left\{\frac{\bbE\big[\phi_n (\bX_0^{n-1})\log\,f_n(\bX_0^{n-1})\big]}{\bbE\big[\phi^*_n (\bX )
\log\,f_n(\bX_0^{n-1})\big]}\right\}^{1/n}\to 1,\hbox{ or }\frac{1}{n}
\log\,\frac{\bbE\big[\phi_n (\bX_0^{n-1})\log\,f_n(\bX_0^{n-1})\big]}{\bbE\big[\phi^*_n (\bX )
\log\,f_n(\bX_0^{n-1})\big]}\to 0.\ee
The main results of this paper can be described as follows. \vskip .5 truecm

(A) For additive or asymptotically additive WFs (i.e., under assumption \eqref{eq:1.4} or \eqref{eq:genAdd})
we analyse the limits
\be\label{eq:rateAdd}{\rm{(i)}}\;\;\rA_0=\lim\limits_{n\to\infty}
\frac{I^\rw_{\phi_n}(\BX_0^{n-1})}{n^2},\;\;\;{\rm{(ii)}}\;\;\rA_0=
\lim_{n\to\infty}\frac{H^\rw_{\phi_n}(f_n)}{n^2}.\ee
%As we said, under condition  \eqref{eq:genAdd3}, we will have \eqref{eq:genAdd2}.
% with $\rA_1=\bbE\vtheta^*(\BX)$.

(B) For multiplicative or asymptotically multiplicative WFs (i.e., under assumptions \eqref{eq:1.4}
or \eqref{eq:genMul}), the focus will be on convergences
\be\label{rhom}
{\rm{(i)}}\;\;\orB_0 =\lim\limits_{n\to\infty}\frac{1}{n}\log\,I^\rw_{\phi_n}(\BX_0^{n-1}),\;\;
{\rm{(ii)}}\;\;\rB_0 =\diy\lim_{n\to\infty}\frac{1}{n}\log\,H^{\rm w}_{\phi_n^{\,}}(f_n).
\ee
%{\rm{(iii)}}\;\;\diy\lim_{n\to\infty}\frac{h^{\rm w}_{\phi_n^{\,}}(\BX_0^{n-1})}{n\,e^{\mu (n-1)}}=\theta,\;\;
%\lim_{n\to\infty}\frac{I^\rw_{\phi_n}(\BX_0^{n-1})}{n\,e^{\eta n}}=\varpi\;\hbox{($\bbP$-a.s.).}\ena \ee
In (\ref{eq:rateAdd}i), (\ref{rhom}i) we bear in mind various forms of convergence for
random variables (see specific statements below).  For multiplicative
WFs  we will also identify an analog of the value $\rB_1$ from \eqref{eq:M1} for Markov chains:
\be\label{rhom3}\rB_1=\lim\limits_{n\to\infty}\frac{H^\rw_{\phi_n}(f_n)}{n\rB_0^{n-1}}.
\ee
We want to stress that some  properties are established
in this paper
under rather restrictive assumptions, although in our opinion, a natural class of RPs for which
these properties hold is much wider. This view is partially supported by an analysis of Gaussian
processes $\BX_0^\infty$ is conducted in Section 5.

\br\label{Remark 2.1.} {\rm The normalisation considered in \eqref{eq:genAdd2}, \eqref{eq:rateAdd}
and \eqref{rhom} is connected with stationarity/ergodicity of RP $\BX$ and various forms of asymptotic
additivity and multiplicativity of WFs $\phi_n$. Abandoning these types of assumptions may
lead to different types of scaling.}
\er

\section{Rates for additive WFs}

%{\bf 3.A. A general statement.}
\subsection{A general statement}\label{3.A}
Consider first a general case where $\BX$ is a stationary
ergodic RP with a probability distribution $\bbP$ on $\bcX$. In this case we write
\be\label{eq:WIGE}\beal I^\rw_{\phi_n}(\bx_0^{n-1})=-\phi_n(\bx_0^{n-1})\left[\log\,p_0(x_0)
+\sum\limits_{1\leq  j< n}\log p^{(j)}(x_j|\bx_0^{j-1})\right].\ena\ee
As in Eqn \eqref{eq:genAdd3}, $p^{(j)}(y|\bx_0^{j-1})$ represents the conditional PM/DF of
having $X_j=y$ given that string $\BX_0^{j-1}$ coincides with $\bx_0^{j-1}$, and $p_0(y)$ is the
PM/DF for $X_0$:
\be\label{eq:3.2}p_0(y)=\frac{\bbP (X_0\in\rd y)}{\nu (\rd y)},\;\;p^{(j)}(y|\bx_0^{j-1})=\frac{\bbP (X_j\in\rd x)}{\nu (\rd y)},
\;\;y\in\cX,\;\bx_0^{j-1}\in\cX^j.\ee
The SE rate $h$ is defined by
\be\label{eq:hGEnrg} h=-\bbE\log\,p(X_0|\BX_{-\infty}^{-1})\ee
where $p(y|\bx_{-\infty}^{-1})$ is the conditional PM/DF for $X_0=y$ given $\bx_{-\infty}^{-1}$,
an infinite past realization of $\BX$.
% (The sequence $H(X_0|\BX_{-n}^{-1})$ is non-decreasing
% with $n$.)
As before, set $H^\rw_{\phi_n}(f_n)=\bbE I^\rw_{\phi_n}(\BX_0^{n-1})$. Recall, the SMB theorem
asserts that for an ergodic RP $\BX$, the following limit exists $\bbP$-a.s.
\be\label{eq:SMBGen1}\lim\limits_{n\to\infty}\diy\frac{1}{n}\left[\log\,p_0(X_0)
+\sum\limits_{1\leq  j< n}\log p^{(j)}(X_j|\bX_0^{j-1})\right]=h.\ee

\bt\label{Thm1}%{\bf Theorem 1.}
Given an ergodic probability distribution $\bbP$ on $\bcX$, consider
the WI $I^\rw_{\phi_n}(\BX_0^{n-1})$ and the WE $H^\rw_{\phi_n}(f_n)$
as defined in \eqref{eq:1.3b} and \eqref{eq:1.3a}. Suppose that convergence
in \eqref{eq:genAdd} holds $\bbP$-a.s. Then:
\begin{description}
  \item[{\rm{(I)}}] Convergence in {\rm{(\ref{eq:rateAdd} i)}} holds
true, $\bbP$-a.s., with $\rA_0=\alpha h$ where $\alpha$ is as in  \eqref{eq:genAdd}
and $h$ as in \eqref{eq:hGEnrg}. That is:
$$\lim\limits_{n\to\infty}
\frac{I^\rw_{\phi_n}(\BX_0^{n-1})}{n^2}=\alpha h,\;\;\bbP\hbox{-a.s.}$$
  \item[{\rm{(II)}}] Furthermore, {\rm{(a)}} suppose that the
WFs $\phi_n$ exhibit convergence \eqref{eq:genAdd}, $\bbP$-a.s.,
with a finite $\alpha$, and $\big|\phi_n(\BX_0^{n-1})/n\big|\leq c$ where $c$ is a
constant independent of $n$. Suppose also that convergence in Eqn \eqref{eq:SMBGen}
holds with $h\in [0,\infty )$ given by \eqref{eq:hGEnrg}.
Then convergence in {\rm{(\ref{eq:rateAdd} ii)}} holds true, as before with $\rA_0=\alpha h$:
$$\lim_{n\to\infty}\frac{H^\rw_{\phi_n}(f_n)}{n^2}=\alpha h.$$
{\rm{(b)}} Likewise, convergence in Eqn {\rm{(\ref{eq:rateAdd} ii)}} holds true whenever
 convergences \eqref{eq:genAdd} and \eqref{eq:SMBGen} hold \ $\bbP$-a.s. and
$|\log\,f_n(\BX_0^{n-1})/n|\leq c$ where $c$ is a constant.
Finally, {\rm{(c)}} suppose that convergence in \eqref{eq:genAdd} and
\eqref{eq:SMBGen} holds in $\rL_2$, with finite $\alpha$ and $h$. Then
convergence in {\rm{(\ref{eq:rateAdd} ii)}} holds true, again with $\rA_0=\alpha h$.
\end{description}
\et

\bp  Assertion (I) follows immediately from the $\bbP$-a.s.
convergence in Eqns \eqref{eq:genAdd}
and \eqref{eq:hGEnrg}. The same is true of assertions (IIa) and (IIb),
with the help of the Lebesgue dominated convergence theorem. Assertion (IIc) follows from
the $\rL_2$-convergence and continuity of the scalar product.
\ep

\br\label{Remark 3.1.} {\rm The assumption in statement (IIc) of Theorem \ref{Thm1}
that the limit in \eqref{eq:SMBGen} holds in $\rL_2$ (i.e., an $\rL_2$-SMB theorem)
can be checked in a number of special cases. We conjecture that a sufficient condition
is that $\bbP$ is ergodic and RV $\log\,p(X_0|\BX_{-\infty}^{-1})$ lies in $\rL_2$. However,
to the best of our knowledge, it is an open question. The fact that the limits in parts (I)
and (IIa) coincide can be considered as an analog of the SMB theorem to the case under
consideration.}
% (Here and below, $p(X_0|\BX_{-\infty}^{-1})$
% stands for the conditional probability given an infinite past generated by $\bbP$.)
\er

\br\label{Remark 3.2.} {\rm Under conditions of Theorem \ref{Thm1}, the bound
$|\log\,f_n(\BX_0^{n-1})/n|\leq c$
in assertion (b) holds when $\cX$ is a finite or a countable set  (the Chung-Neveu lemma).}
\er

\br\label{Remark 3.3.} {\rm The factor \ $\diy\frac{1}{n}$ \ in assumption \eqref{eq:genAdd} can
be replaced by
$\diy\frac{1}{a(n)}$ where $a(n)$ is a given increasing sequence of positive numbers.
In this case we can speak of a moderated asymptotic additivity of WF $\phi_n$.
Accordingly, in (\ref{eq:rateAdd}) the denominator $n^2$ should be replaced with $na(n)$.}
\er

\br\label{Remark 3.4.} {\rm The statement of Theorem \ref{Thm1} remains in force when
in representation
\eqref{eq:WIGE} the sum $\log\,p_0(X_0)
+\sum\limits_{1\leq  j< n}\log p^{(j)}(X_j|\bX_0^{j-1})$ is replaced with $\sum\limits_{j=0}^{n-1}
p(X_j|\BX_{-\infty}^{j-1})$ and/or WF $\phi_n(\BX_0^{n-1})$
is replaced by the sum $ \phi^*_n(\BX)=\sum\limits_{j=0}^{n-1}\vphi^*(S^j\BX)$ (cf. Eqn
\eqref{CondB}), under appropriate assumptions upon $\vphi^*$. This is achieved by making
use of standard Ergodic theorems (Birkhoff and von Neumann).
}
\er

\subsection{The Markovian case}\label{3.B. The Markovian case}

It is instructive to affiliate an assertion analogous to Theorem \ref{Thm1} for a Markov chain of
order $k\geq 1$. In this case the PM/DF $f_n(\bx_0^{n-1})$, relative to reference
measure $\nu^k$ on $\cX^k$, for $n>k$ has the form
\be\label{eq:lamknPDF}\beal f_n(\bx_0^{n-1})=\lam (\bx_0^{k-1})\prod\limits_{0\leq  j< n-k}
p(x_{j+k}|\bx_j^{j+k-1}).\ena\ee
Here $\lam$ yields a PM/DF for an initial string: $\lam (\bx_0^{k-1})\geq 0$
and $\int\limits_{\cX^k}\lam (\bx_0^{k-1})\nu^k(\rd\bx_0^{k-1})=1$. Further, as
above, $p(y|\bx_j^{j+k-1})$ represents the conditional PM/DF of having $X_{j+k}=y$ given that
string $\BX_j^{j+k-1}$ coincides with $\bx_j^{j+k-1}$. Next, let $\pi$ be an equilibrium PM/DF on
$\cX^k$, with
\be\label{eq:equipi}\beal
\pi (\bx_0^{k-1})=\int\limits_\cX\pi (x'\vee \bx_0^{k-2})p(x_{k-1}|x'\vee \bx_0^{k-2})
\nu (\rd x')\ena\ee
where string $x'\vee\bx_0^{k-1}=(x',x_0,\ldots ,x_{k-2})\in\cX^k$.
Denote by $\bbP_\lam$
and $\bbP=\bbP_\pi$ the probability distributions (on ${\bcX}_+$
and $\bcX$, respectively) generated by the process with initial PM/DF $\lam$ and $\pi$.
Further, let $\bbE$ and $\bbE_\lam$ stand for the expectations under $\bbP$ and $\bbP_\lam$.
Set
\be\label{eq:hEntMCk}\beal h=-\bbE\log\,p(X_k|\bX_0^{k-1})= -\int\limits_{\cX^k}\pi (\bx_0^{k-1})
p^{(k)}(x_k|\bx_0^{k-1})\log\,p^{(k)}(x_k|\bx_0^{k-1})\nu^k(\rd\bx_0^k).\ena\ee
Next, define $H^{\rm w}_{\phi_n}(f_n,\pi )=\bbE I^{\rm w}_{\phi_n}(\BX_0^{n-1})$ and
$H^{\rm w}_{\phi_n}(f_n,\lam)=\bbE_\lam I^{\rm w}_{\phi_n}(\BX_0^{n-1},\lam )$ where
\be\label{eq:WIMakG}\beal I^\rw_{\phi_n}(\bx_0^{n-1},\lam )=-\phi_n(\bx_0^{n-1})\left[\log\lam (\bx_0^{k-1})
+\sum\limits_{0\leq  j< n-k}\log p^{(j+k)}(x_{j+k}|\bx_j^{j+k-1})\right].\ena\ee

For definiteness, in Theorem 3.2 below we adopt conditions in a rather strong form, without
distinguishing between different possibilities listed in the body of Theorem 3.1. The proof
of Theorem 3.2 is essentially a repetition of that of Theorem 3.1, with an additional help
from the Ergodic theorems.

\bt\label{Theorem 3.2.}
Let $\BX_0^\infty$ be a $k$-order Markov chain with an initial PM/DF
$\lam (\bx_0^{k-1})$ where
$k\geq 1$. Assume that {\rm{(i)}} Eqn \eqref{eq:genAdd} is
fulfilled, both in $\rL_2$ and $\bbP$-a.s., {\rm{(ii)}} the stationary probability measure $\bbP$
on $\bcX$ is ergodic,
{\rm{(iii)}} $\log\lam (\BX_0^{k-1})$ and $\log\,p (X_k|\BX_0^{k-1})$ belong to $\rL_2$,
{\rm{(iv)}} ${\rm{supp}}\,\lam\subseteq{\rm{supp}}\,\pi$.  Then
the limiting relations \eqref{eq:rateAdd} are satisfied, for both choices of $I^\rw_{\phi_n}(\bx_0^{n-1})$,
$H^\rw_{\phi_n}(f_n)$ and of $I^\rw_{\phi_n}(\bx_0^{n-1},\lam )$, $H^\rw_{\phi_n}(f_n,\lam )$,
with $A=\alpha h$ where $\alpha$ is as in  \eqref{eq:genAdd} and $h$ as in \eqref{eq:hEntMCk}.
Correspondingly, convergence in {\rm{(\ref{eq:rateAdd} i)}} holds $\bbP$-a.s. and $\bbP_\lam$-a.s.
% Convergence in {\rm{(\ref{eq:rateAdd} ii)}} takes place for both $H^\rw_{\phi_n}(f_n)$ and
% $H^\rw_{\phi_n}(f_n,\lam)$.}
\et

A similar assertion could be given in the case of a general initial probability distribution
$\blam (\rd\bx_0^{k-1})$ on
$\cX^k$ which can be singular relative to $\nu^k$. Here, for $n>k$ we
consider the PM/DF $f_n(\bx_0^{n-1})$ with respect to
$\blam (\rd\bx_0^{k-1})\nu^{n-k}(\rd\bx_k^{n-1})$ on $\cX^n$:
\be\label{eq:3.8}\beal f_n(\bx_0^{n-1})=\prod\limits_{1\leq j< n}p(x_j|\bx_{j-k}^{j-1}),\;\hbox{ with
$\int\limits_{\cX^n}f_n(\bx_0^{n-1})\blam (\rd x_0)
\nu^{n-k}(\rd\bx_k^{n-1})=1$.}\ena\ee
Then $\bbP_\blam$ denotes the probability distribution (on $\bcX_+$) generated
by the process with the initial distribution $\blam$ whereas $\bbE_\blam$ stands for the expectation
under $\bbP_\blam$. The notation $\bbP=\bbP_\pi$ and $\bbE=\bbE_\pi$ has the same meaning
as before, with $\pi(\bx_0^{k-1})$ being an equilibrium PM/DF relative to $\nu^k$ on $\cX^k$.
Accordingly, we now define
\be\label{eq:WIMakGG}\beal I^\rw_{\phi_n}(\bx_0^{n-1})=-\phi_n(\bx_0^{n-1})
\sum\limits_{0\leq  j< n-k}\log p^{(j+k)}(x_{j+k}|\bx_j^{j+k-1})\ena\ee
and $H^{\rm w}_{\phi_n}(f_n,\blam)=\bbE_\blam I^{\rm w}_{\phi_n}(\BX_0^{n-1})$.

\bt\label{Theorem 3.3.}
Let $\BX_0^\infty$ be a $k$-order Markov chain with an initial probability
measure $\blam (\rd\bx_0^{k-1})$ where
$k\geq 1$. Adopt assumptions {\rm{(i)}} and {\rm{(ii)}} of Theorem \ $3.2$. In addition, suppose that
{\rm{(iii)}} $p^{(k)}X_k|\BX_0^{k-1})>0$ \ $\bbP$-a.s. (implying that $\pi (\bx_0^{k-1})$ is strictly
positive $\nu^k$ -a.s. on $\cX^k$)
and that $\log\,p^{(k)}(X_k|\BX_0^{k-1})$ belongs to $\rL_2$.
With $I^\rw_{\phi_n}(\bx_0^{n-1})$ as in \eqref{eq:WIMakGG},
the assertions of Theorem \ $3.2$ hold true, mutatis mutandis, and
convergence in {\rm{(\ref{eq:rateAdd} i)}} takes place $\bbP$-a.s. and $\bbP_\blam$-a.s. Furthermore,
convergence in {\rm{(\ref{eq:rateAdd} ii)}} holds for both $H^{\rm w}_{\phi_n}(f_n)$ and
$H^{\rm w}_{\phi_n}(f_n,\blam)$.
\et

\bt\label{Theorem 3.4.}
Suppose that $\phi_n(\bx_0^{n-1})=\sum\limits_{j=0}^{n-1}\vphi (x_j)$. Let
$\BX$ be a stationary RP with the property that $\forall$ $i\in\bbZ$ there exists the limit
\be\label{eq:rA1id}\bear -\rA_1:=\lim\limits_{n\to\infty}\sum\limits_{j\in\bbZ:\,|j+i|\leq n}
\bbE\big[\vphi (X_0)\log p^{(n+i+j)}(X_j|\BX_{-n-i}^{j-1})\big]
\qquad\qquad{}\\
=\sum\limits_{j\in\bbZ}\bbE\big[\vphi (X_0)\log p(X_j|\BX_{-\infty}^{j-1})\big],\ena\ee
and the last series converges absolutely. Then
$\lim\limits_{n\to\infty}{\diy\frac{1}{n}}H^\rw_{\phi_n}(f_n)=\rA_1$.
\et

\bp Set: $n_1=-[n/2]$, $n_2=[(n+1)/2]-1$. Then we can write
\be\label{eq:WIGE2}\beal {\diy\frac{1}{n}}H^\rw_{\phi_n}(f_n)=-{\diy\frac{1}{n}}
\sum\limits_{n_1\leq i\leq n_2}\rE_{n,i}\\
\qquad\qquad\hbox{where }\;\rE_{n,i}=\sum\limits_{l=n_1-i}^{n_2-i}
\bbE\big[\vphi (X_0)\log p^{(-n_1+l+i)}(X_{l}|\BX_{n_1-i}^{l-1})\big].\ena\ee
(For $l=n_1-i$, we have the term $\log\,p_0(X_{n_1-i})$.)
By virtue of \eqref{eq:rA1id}, each $\rE_{n,i}$ tends to $-\rA_1$, hence the Cesaro mean
does too.
\ep

\br\label{Remark 3.5.} {\rm Condition \eqref{eq:rA1id} alludes that $\bbE\vphi (X_i)=0$. We will now show
that \eqref{eq:rA1id} holds when $\cX$ is a finite set and $\BX$ is a stationary ergodic Markov chain
with positive transition probabilities $\tp (x,y)$ and equilibrium probabilities $\pi (x)$, $x,y\in\cX$.
Then $\rho :=\min\;\tp (x,y)$ satisfies $0<\rho <1$, and the $s$-step transition probabilities
$\tp^{(s)}(x,y)$ obey $|\tp^{(s)}(x,y)-\pi (y)|\leq 2(1-\rho )^s$ (a Doeblin property).  Assume that
$\bbE\vphi (X_i)=
\sum\limits_{x\in\cX}\pi (x)\vphi (x)=0$. Then, $\forall$ $n_1,n_2\in\bbZ$ with $n_1<0<n_2$,
\be\label{eq:WIGE3}\beal \sum\limits_{j\in\bbZ:\,n_1\leq j\leq n_2}
\bbE\big[\vphi (X_0)\log p_{j-n_1}(X_j|\BX_{n_1}^{j-1})\big]\\
\qquad = \sum\limits_{n_1\leq j\leq 0}\sum\limits_{x,y,z\in\cX}
\big[\pi (x)\tp(x,y)\log\tp(x,y)\tp^{(-j)}(y,z)\vphi (z)\big]\\
\qquad\qquad +\sum\limits_{0\leq j<n_2}\sum\limits_{x,y,z\in\cX}
\big[\vphi (x)\pi (x)\tp^{(j-1)}(x,y)\tp(y,z)\log\tp(y,z)\big].\ena\ee
As $-n_1,n_2\to\infty$, the RHS in \eqref{eq:WIGE3} represents absolutely convergent series; this
leads to \eqref{eq:rA1id}.}
\er

\br\label{Remark 3.6.} {\rm Condition \eqref{eq:rA1id} is equivalent to the condition of combined
asymptotic expected additivity from \eqref{eq:genAdd3}.}
\er

\subsection{The Gaussian case}\label{3.C. The Gaussian case}

Gaussian processes (GPs) form an instructive example casting
light upon the structure of the primary WE rate $\rA_0$: they give an opportunity to assess
an impact of ergdicity and asymptotic additivity. Here we list and discuss GP properties
in a convenient order.
Consider a real double-infinite matrix $\tC =(C(i,j):\;i,j\in\bbZ)$. Assume that,
$\forall$ $m<n$, the $(n-m+1)\times (n-m+1)$ bloc $\tC_{m,n}=(C(i,j):\,m\leq i,j\leq n)$ gives
a (strictly) positive definite matrix. A GP $\BX =(X_n:\;n\in\bbZ)$ with zero mean and covariance
matrix $\tC$ has a family of PDFs $f_{m,n}=f^{\rm{No}}_{\diy\tC_{m,n}}$, $m<n$, in $\bbR^{n-m+1}$,
relative to the Lebesgue measure \ $\rd\bx_{m,n}$. Here
\beq\label{eq:GaussPDF}\beacl f_{m,n}(\bx_{m,n})&={\diy\frac{1}{\Big[(2\pi)^{n-m+1}{\rm{det}}\,
\tC_{m,n} \Big]^{1/2}}}
\exp\left(-{\diy\frac{\bx_{m,n}^\rT\tC_{m,n}^{-1}\bx_{m,n}}{2}}\right),\\
\;\;&\qquad\qquad\qquad\qquad\qquad\bx_{m,n}^\rT=\bx_m^n=(x_m,\ldots ,x_n)\in\bbR^{n-m+1}.
\ena\eeq
In this section, $\bx_{m,n}$ stands for a column-
and $\bx^\rT_{m,n}$ for a row-vector. (A similar rule will be applied to random vectors
$\BX_{m,n}$ and $\BX_{m,n}^\rT$.) When $m=0$ we write
$f_n$ for $f_{0,n}$ and $\tC_n$ for $\tC_{0,n}$.

If entries $C(i,j)$
have the property $C(i,j)=C(0,j-i)$,  process $\BX$ is stationary.
In this case the spectral measure is a (positive) measure $\mu$ on $[-\pi, \pi )$ such that
$C(i,j)=\int\limits_{-\pi}^\pi\cos [(j-i)s]\mu (\rd s)$.
A stationary GP $\BX$ is ergodic iff $\mu$ has no atoms. Various forms of
regularity (decay of correlation) of GPs have been presented in great detail in
\cite{IR}.We want to note that in Theoretical and Applied Probability (as well as in Statistics),
the basic parameter is, typically, $\tC$. On the other hand,
in Mathematical Physics it is usually the family of matrices $\tC^{-1}_{m,n}$: their entries
$C^{(-1)}_{m,n}(i,j)$ play the role of interaction potentials between sites $m\leq i,j\leq n$ for a
system of `spins' $x_m,\ldots ,x_n\in\bbR$. In this interpretation, the quadratic form \ $\diy\frac{1}{2}\bx_{m,n}^\rT\tC_{m,n}^{-1}\bx_{m,n}$ \
represents the potential energy of a spin configuration $\bx_{m,n}$.
In these terms, a Markov GP arises when matrices $\tC^{-1}_{m,n}$ are tri-diagonal Jacobi;
cf. Eqn \eqref{eq:GssMrkvfn} below. The SE $H(f_{m,n})=\diy\frac{1}{2}\log \left[ e(2\pi)^n ({\rm{det}}\,
\tC_n)\right]={\diy\frac{1}{2}}\Big[n\log\,(2\pi e)-{\rm{tr}}\;\tL_n\Big]$ where $\tL_n=\log\,\tC_n^{-1}$.

Now take $m=0$. Given a WF
$\bx_{0,n-1}\in\bbR^n\mapsto\phi_n(\bx_{0,n-1})$, the WI and WE have the form
\be\label{eq:WIGau}
\beal I^\rw_{\phi_n} (\bx_0^{n-1})={\diy\frac{\log \left[ (2\pi)^n ({\rm{det}}\,
\tC_n)\right]}{2}}
\phi_n (\bx_{0,n-1})+{\diy\frac{\log\,e}{2}}\left(\bx_{0,n-1}^\rT\tC_n^{-1}
\bx_{0,n-1}\right) \phi_n (\bx_{0,n-1})\ena \ee
and
\be\label{eq:WEGau}\beal H^\rw_{\phi_n} (f_n)={\diy\frac{1}{2}} \log \left[ (2\pi)^n ({\rm{det}}\,
\tC_n)\right]\int\limits_{\bbR^n}
\phi_n (\bx_{0,n-1})f_n (\bx_{0,n-1})\rd\bx_{0,n-1}\\
\qquad\qquad\quad +{\diy\frac{\log\,e}{2}}\int\limits_{\bbR^n}\left(\bx_{0,n-1}^\rT\tC_n^{-1}
\bx_{0,n-1}\right) \phi_n (\bx_{0,n-1})f_n (\bx_{0,n-1})\rd\bx_{0,n-1}\\
\quad =\left[H(f_n)-n{\diy\frac{\log\,\re}{2}}\right]\bbE\phi_n(\BX_{0,n-1})
+{\diy\frac{\log\,e}{2}}
\bbE\Big[\left(\BX_{0,n-1}^\rT\tC^{-1}_n\BX_{0,n-1}\right)\phi_n(\BX_{0,n-1})
\Big]. \ena \ee
Consequently, a finite rate $h=\lim\limits_{n\to\infty}{\diy\frac{1}{n}}H (f_n)$ exists
iff
\be\label{eq:GaussL}\beal
\lim\limits_{n\to\infty}{\diy\frac{1}{n}}{\rm{tr}}\;\tL_n=-h+\log\,(2\pi\re),\ena\ee
regardless of ergodicity (and even stationarity) of GP $\BX$. Moreover,
under assumption \eqref{eq:GaussL}, we obtain that
\be\label{eq:GaussGg}\frac{H^\rw_{\phi_n}(f_n)-(\log\,e)
\bbE\big[\left(\BX_{0,n-1}^\rT\tC^{-1}_n\BX_{0,n-1}\right)
\phi_n(\BX_{0,n-1})\big]/2}{n\bbE\phi_n(\BX_{0,n-1})}\to h-\frac{\log\re}{2}\ee
for any choice of the WFs $\phi_n$ such that $\bbE\phi_n(\BX_{0,n-1})\neq 0$.
For an asymptotically additive WF $\phi_n$ satisfying \eqref{eq:genAdd} and for a GP
obeying \eqref{eq:GaussL}, Eqn \eqref{eq:GaussGg} takes the form
$$\frac{H^\rw_{\phi_n}(f_n)-(\log\,e)\alpha n^2/2}{\alpha n^2}\to h-\frac{\log\re}{2} .$$
This yields (\ref{eq:rateAdd} i) with $\rA_0=\alpha h$, again without
using ergodicity/stationarity of $\BX_0^\infty$.

Similarly, \eqref{eq:WIGau} and \eqref{eq:GaussL} imply that $\forall$ $\bx\in\bcX$,
\be\label{eq:GaussgG}\frac{I^\rw_{\phi_n}(\bx_0^{n-1})-(\log\,e)
\left(\bx_{0,n-1}^\rT\tC^{-1}_n\bx_{0,n-1}\right)
\phi_n(\bx_{0,n-1})/2}{\phi_n(\bx_{0,n-1})}\to h-\frac{\log\re}{2}\ee
for any choice of the WFs $\phi_n$ such that $\phi_n(\bx_{0,n-1})\neq 0$.

On the other hand, take $\phi_n(\bx_{0,n-1})=\alpha n$ (an additive WF with $\vphi (x)=\alpha$).
Then Eqn \eqref{eq:WEGau} becomes
\be\label{eq:WEGaubn}\beal H^\rw_{\phi_n} (f_n)={\diy\frac{\alpha n}{2}}\Big[n\log (2\pi e)-
{\rm{tr}}\;\tL_n\Big]=\alpha nH(f_n).\ena\ee
The asymptotics for the WE $H^\rw_{\phi_n} (f_n)$ and SE
$H (f_n)$ will be
determined by a `competition' between the terms in the square brackets
(an entropy-energy argument in Mathematical Physics). Viz., take
$\tL_n=(L_n(i,j)$, $0\leq i,j<n)$ and suppose that the diagonal
entries decrease to $-\infty$ when $j$ is large (say, $L(j,j)\sim -\log (c+j)$ with
a constant $c>0$ or $\lam_j\sim e^{(c+j)}$ where
$\lam_0\leq \lam_1\leq\ldots\leq\lam_{n-1}$ are the eigen-values of $\tC_n$). Then
the trace ${\rm{tr}}\;\tL_n =\sum\limits_{0\leq j<n}L_n(j,j)$ will dominate, and the
correct scale for the rate of $H^\rw_{\phi_n} (f_n)$ with $\phi_n(\bx_{0,n-1})=\alpha n$
will be $\;\diy\frac{1}{n^2\log\,n}$.

The above example can be generalised as follows.
% In the context of asymptotic additivity, a natural family of WFs $\bx_{m,n}\in\bbR^{n-m+1}
% \mapsto\phi_{m,n} (\bx_{m,n})$ is of the following form.
Let $\tA =(A(i,j):\;i,j\in\bbZ)$ be a double-infinite real symmetric matrix
(with $A(i,j)=A(j,i)$) and consider, $\forall$ $m<n$,
the bloc $\tA_{m,n}=(A(i,j):\,m\leq i,j\leq n)$. Then set
% is such that matrix $\tC_{m,n}^{-1}-\tA_{m,n}$ is (strictly) positive definite.
% choose a real double-infinite sequence $\bbt =(t_n,\;n\in\bbZ)$ and
\beq\phi_{m,n} (\bx_{m,n})=\bx_{m,n}^\rT\tA_{m,n}\bx_{m,n}.\eeq
% \bx_{m,n}^\rT\big(\tC_{m,n}^{-1}-\tA_{m,n}\big)\bbt_{m,n}+
% where $\bbt_{m,n}=(t_i:m\leq i\leq n)$.
For $\tA_{0,n-1}$ we write $\tA_n$.
Pictorially, we try to combine a Gaussian form
of the PDFs $f_{m,n}(\bx_{m,n})\;$ with a log-Gaussian form of $\phi_{m,n}(\bx_{m,n})$.

Then the expression for the WI $\;I^\rw_{\phi_n}(\bx_{0,n-1})=-\phi_n(\bx_{0,n-1})
\log\,f_n(\bx_{0,n-1})\;$ and WE $H^\rw_{\phi_n}(f_n)=\bbE I^\rw_{\phi_n}(\BX_{0,n-1})$
become
\beq\label{eq:WIGaGa}\beal I^\rw_{\phi_n}(\bx_{0,n-1})=\left(\bx_{0,n-1}^\rT
\tA_n\bx_{0,n-1}\right)\\
\qquad\qquad\quad\quad\times\left\{
\left[H(f_n)-n{\diy\frac{\log\re}{2}}\right]+\Big(\bx_{0,n-1}^\rT\tC_n^{-1}\bx_{0,n-1}\Big)\log\,\re\right\}
% \qquad\qquad\\ \times \big[\big]\left[\bx_{0,n-1}^\rT\Big(\tC_n^{-1}-\tA_n\Big)\bbt_{0,n-1}
\ena\eeq
and
\beq\label{eq:WEGaGa}\bear H^\rw_{\phi_n}(f_n)=\left[H(f_n)-n{\diy\frac{\log\re}{2}}\right]
\bbE\Big(\BX_{0,n-1}^\rT\tA_n\BX_{0,n-1}\Big)\qquad\qquad\qquad\\
 +\diy\frac{\log\,\re}{2}\bbE\Big[\Big(\BX_{0,n-1}^\rT\tC_n^{-1}\BX_{0,n-1}\Big)
\Big(
%\BX_{0,n-1}^\rT\Big(\tC_n^{-1}-\tA_n\Big)\bbt_{0,n-1} +{\diy\frac{1}{2}}
\BX_{0,n-1}^\rT\tA_n\BX_{0,n-1}\Big)\Big].\ena\eeq

As before, the analysis of rates for \eqref{eq:WIGaGa} and \eqref{eq:WEGaGa} can be done
by comparing the contributions from different terms.

\section{Rates for multiplicative WFs}

Multiplicative weighted rates behave differently and require a diverse approach to their
studies. To start with, the WI rate in general does not coincide with the corresponding
WE rate.

\subsection{WI rates}\label{4.A. WI rates.}

The question of
a multiplicative WI rate is relatively simple:

\bt\label{ Theorem 5.}
Given an ergodic RP $\BX$ with a probability distribution $\bbP$ on
$\bcX$, consider the WI $I^\rw_{\phi_n}(\bx_0^{n-1})$
as defined in \eqref{eq:WIGeIH} and \eqref{eq:WIGE}. Suppose that convergence
in \eqref{eq:genMul} holds $\bbP$-a.s. Then convergence in {\rm{(\ref{rhom} i)}} holds
true $\bbP$-a.s., where $\orB=\beta$ and the value $\beta$ is as in \eqref{eq:genMul}.
\et

\bp The assertion follows immediately from the $\bbP$-a.s.
convergence in Eqn \eqref{eq:genMul}.
\ep

\subsection{WE rates. The Markovian case}\label{4.B. WE rates. The
Markovian case}

Passing to multiplicative WE rates, we consider
in this paper a relatively simple case where (a) RP $\BX_0^\infty$ is  a homogeneous
MC with a stationary PM/DF $\pi (x)$ and
the conditional  PM/DF $p(y|x)$ and (b) the WF $\phi_n(\bx_0^{n-1})$  is a product:
for $x,y\in\cX$ and $\bx_0^{n-1}=(x_0,\ldots ,x_{n-1})\in\cX^n$,
\be\label{eq:condip}\beal p(y|x)=\diy\frac{\bbP(X_k\in\rd y|X_{k-1}=x)}{\nu (\rd y)},
\;\;\phi_n(\bx_0^{n-1})=\prod\limits_{0\leq j<n}\vphi (x_j).\ena\ee
In this sub-section we assume that $\vphi (x)\geq 0$ on $\cX$ and adopt some
positivity assumptions on $p(y|x)$: there exists $k\geq 0$ such that
\beq \label{eq:Doebl}p^{(k+1)}(y|x)=\int_{\cX^{k}} p(y|u_{k})\cdots
p(u_1|x)\nu^{k}(\rd\bu_1^{k})>0.\eeq
As earlier, $\lam $ stands for an initial PM/DF on $\cX$. Accordingly, we consider
the WE $H^\rw_{\phi_n}(f_n,\lam )$ of the form
\be\label{eq:WEkMC}\beal H^\rw_{\phi_n}(f_n,\lam )\;= -\bbE_\lam\left\{
\prod\limits_{0\leq j< n}
\vphi (X_j)\log\left[\lam (X_0)\prod\limits_{1\leq l< n}
p(X_l|X_{l-1})\right]\right\}\\
\qquad\qquad\quad =-\int_{\cX^n}\lam (x_0)\vphi (x_0)\prod\limits_{1\leq i< n}
\big[p(x_i|x_{i-1})\vphi (x_i)\big]\\
\qquad\qquad\qquad\qquad\times
\left[\log\lam (x_0)+\sum\limits_{1\leq l< n}
\log p(x_l|x_{l-1})\right]\nu^n(\rd\bx_0^{n-1}),\ena\ee
and the WE $H^\rw_{\phi_n}(f_n)$ obtained by replacing $\lam$ with $\pi$.

The product $\vphi (x_0)\prod\limits_{1\leq i< n}
\big[p(x_i|x_{i-1})\vphi (x_i)\big]$ can of course be written in a symmetric (or dual)
formation, as $\prod\limits_{1\leq i< n}
\big[\vphi (x_{i-1})p(x_i|x_{i-1})\big]\vphi (x_{n-1})$. It would lead to an equivalent
form of results that follow.

% In Theorem 6 below we assume that the reference measure $\nu$ on $\cX$ is finite:
% $\nu (\cX)<\infty$. {\bf Na samom dele, eto uslovie ne nuzhno??}
% (This assumption can be removed at a cost of involved technicalities.) Our aim here is to promote
% a suitable methodology, leaving the goal of achieving technically involved results to future works.
% However, see Remark 4.3 below.

The existence (and a number of properties) of the WER $\rB_0$ in (\ref{rhom} ii) are related
to an integral operator $\tW$ acting on functions $\tf:\cX\to\bbR$ and connected to the
conditional PM/DF $p(y|x)$ and factor $\vphi (x)$ in \eqref{eq:condip}.
Namely, for $y\in\cX$, the value $(\tW\tf )(y)$ is defined by
\be\label{eq:opQ}\bear (\tW\tf)(y)=\int_\cX W(y,w)\tf(w)\nu (\rd w).\ena\ee
We also introduce an adjoint/transposed operator $\tW^\rT$ with an action $\tg\mapsto\tg\tW^\rT$:
\be\label{eq:opQT}\bear\left(\tg\tW^\rT \right)(y)=\int_{\cX}\tg(w)
W (w,y)\nu (\rd w).\ena\ee
Here the  kernel $W$ given as follows: for $u,v\in\cX$,
\be\label{eq:kenW}W(u,v)=\vphi (u)p (v|u).\ee
% \;\;W^\rT(\bu_1^k;v)=\vphi (\bu_1^k)p (v|\bu_1^{k-1}). \ee
% (As in \eqref{eq:equipi}, symbol $\vee$ stands for concatenations (left and right).)
% $$x\vee\by_1^{k-1}=(x,y_1,\ldots ,y_{k-1})\in\cX^k\;\;\hbox{ and }\;\;
% \by_1^{k-1}\vee x =(y_1,\ldots ,y_{k-1},x)\in\cX^k.$$

\br\label{Remark 4.1.}
{\rm The form of writing the action of the adjoint operator as $\tg\tW^\rT$ does not have
a particular significance but shortens and makes more transparent some relations where $\tW$
and $\tW^\rT$ take part. Viz., we have that
$$\beal\int\limits_{\cX}\tg(y)(\tW\tf)(y)\nu (\rd y)=\int\limits_{\cX}(\tg\tW^\rT)(y)\tf(y)\nu (\rd y),\ena$$
or, in brief, $\left\langle \tg,\tW\tf\right\rangle
=\left\langle \tg\tW^\rT,\tf\right\rangle$
where $\left\langle \tg,\tf\right\rangle =\int\limits_{\cX}\tf (y)\tg(y)\nu (\rd y)$
is the inner product in the (real) Hilbert space $\rL_2(\cX,\nu )$. Also, it emphasizes analogies
with a MC formalism where a transition operator acts on functions while its adjoint (dual) acts on
measures.}
\er

Pictorially speaking,  kernel $W^\rT(x_{i-1};x_i)$ represents the factor
in the product\\ $\prod\limits_{1\leq i< n}\big[p(x_i|x_{i-1})\vphi (x_{i-1})\big]$
in \eqref{eq:WEkMC} where variable $x_i$ appears for the first time.
% (as the rightmost digit in string $\bx_{i-k}^{i}\in\cX^{k+1}$), while
% `moving' along the string $\bx_0^{n-1}\in\cX^n$ from left to right. Similarly,
% $W^\rT(x_{i-k};\bx_{i-k+1}^i;x_{i})$ represents the factor
% where variable $x_{i-k}$ appears for the first time while`moving'  from right to left.
Accordingly:
\be\label{eq:WEkMCW}\beal H^\rw_{\phi_n}(f_n,\lam )=-
\int\limits_{\cX^n}\lam (x_0)\bigg\{\big[\log\lam (x_0)\big]\prod\limits_{1\leq i<n}
W(x_{i-1},x_i)\vphi (x_{n-1})\\
\qquad +\sum\limits_{1\leq l< n}\prod\limits_{1\leq i\leq  l}
W^\rT(x_{i-1}, x_i)\\
\qquad\qquad\quad\times
\left[
\log p(x_l|x_{l-1})\right]\prod\limits_{l<j< n}W(x_{j-1},x_j)
\vphi (x_{n-1})\bigg\}\nu^n(\rd\bx_0^{n-1}).\ena\ee

% Assuming that $p(x_k|\bx_0^{k-1})>0$, $\bx_0^k\in\cX^{k+1}$, the $k$th iterations $W^{(k)}$
% (determining operator $\tW^k$) and ${W^{(k)}}^\rT$ (for operator ${\tW^\rT}^k$)
% become positive kernels:
% \be\label{eq:kernsWk}\beal W^{(k)}(\by_1^k;\bu_1^k)=\vphi (\by_1^k)p (u_1|\by_1^k)
% \cdots\vphi (y_k\vee\bu_1^{k-1})p (u_k|y_k\vee\bu_1^{k-1}),\\
% {W^{(k)}}^\rT(\bu_1^k;\by_1^k)=\vphi (\bu_1^k)p (y_1|\bu_1^k)
% \cdots\vphi (u_k\vee\by_1^{k-1})p (y_k|u_k\vee\by_1^{k-1}),\ena\ee
% with
% \be\label{eq:opsWk}\beac\left(\tW^k\tf\right)(\by_1^k)=\int\limits_{\cX^k}W^{(k)}(\by_1^k;\bu_1^k)
% \tf (\bu_1^k)\nu^k (\rd\bu_1^k),\\
% \left(\tg{\tW^\rT}^k\right)(\by_1^k)=\int\limits_{\cX^k}\tg(\bu_1^k){W^{(k)}}^\rT (\bu_1^k;\by_1^k)
% \nu^k (\rd\bu_1^k),\ena\quad\by_1^k\in\cX^k.\ee

We will use the following condition (of the Hilbert--Schmidt type):
\be\label{eq:condHS}\beal\int\limits_{\cX\times\cX}W(x,y)
W(y,x)\nu (\rd x)\nu (\rd y)<\infty .\ena\ee
Also, suppose that function
\be\label{eq:condBd0} (x,y)\in\cX\times\cX\mapsto p(y|x)|\log p(y|x)|\ee
is bounded and functions
\be\label{eq:condL2}x\mapsto\vphi (x),\;x\mapsto\lam (x)\log\lam (x),\;
x\mapsto\pi (x)\log\pi (x)\ee
belong to $\rL_2(\cX,\nu )$.

\bt\label{Theorem 6.}
Assume the stated conditions upon $\BX_0^\infty$, transitions PM/DF
$p(y|x)$ and WF $\phi_n$.
Then Eqn {\rm{(\ref{rhom} ii)}} holds true, both for  $H^\rw_{\phi_n}(f_n)$
and $H^\rw_{\phi_n}(f_n ;\lam)$:
\beq\lim\limits_{n\to\infty}\frac{1}{n}\log\,H^\rw_{\phi_n}(f_n,\lam )=\lim\limits_{n\to\infty}\frac{1}{n}\log\,H^\rw_{\phi_n}(f_n)=\rB_0.\eeq
Here
\be \rB_0\;=\log\,\mu \ee
and $\mu >0$ is the maximal eigen-value of operator $\;\tW$ in $\rL_2(\cX,\nu)$
coinciding with the norm of \  $\tW$ and  \ $\tW^\rT$; cf. \eqref{eq:opQ}. That is, $\mu =\|\tW\|=\|\tW^\rT\|$.
\et

\bp  As follows from the previous formulas, we
have the following expressions for the WEs $H^\rw_{\phi_n}(f_n,\lam )$ and $H^\rw_{\phi_n}(f_n)$:
\be\label{eq:WEkMCWO}\beal H^\rw_{\phi_n}(f_n,\lam )=-
\int\limits_{\cX}\big[\lam (x_0)\log\lam (x_0)\big]
\left(\tW^{n-1}\vphi\right) (x_{0})\nu (\rd x_0)\\
\quad -\sum\limits_{1\leq l< n}\;\int\limits_{\cX^2}
\left(\lam{\tW^\rT}^{l-1}\right) (x_{l-1})\big[\vphi (x_{l-1})p(x_l|x_{l-1})\\
\qquad\qquad\qquad\qquad\times
\log p(x_l|x_{l-1})\big]\left(\tW^{n-1-l}
\vphi\right) (x_l)\nu^{2}(\rd x_{l-1}\times\rd x_{l})\ena\ee
and
\be\label{eq:WEkMCWP}\beal H^\rw_{\phi_n}(f_n)=-
\int\limits_{\cX}\big[\pi (x_0)\log\pi (x_0)\big]\left(\tW^{n-1}\vphi\right) (x_{0})
\nu (\rd x_0)\\
\quad -\sum\limits_{1\leq l< n}\;\int\limits_{\cX^2}
\left(\pi{\tW^\rT}^{l-1}\right) (x_{l-1})\big[\vphi (x_{l-1})p(x_l|x_{l-1})\\
\qquad\qquad\qquad\qquad\times
\log p(x_l|x_{l-1})\big]\left(\tW^{n-1-l}
\vphi\right) (x_{l})\nu^2(\rd x_{l-1}\times\rd x_{l}).\ena\ee

Re-write \eqref{eq:WEkMCWO} and \eqref{eq:WEkMCWP} by omitting unnecessary
references to $l$:
\be\label{eq:WEkMCWOy}\beal H^\rw_{\phi_n}(f_n,\lam )=
-\int\limits_{\cX}\big[\lam (x)\log\lam (x)\big]
\left(\tW^{n-1}\vphi\right) (x)\nu (\rd x)\\
\quad -\sum\limits_{1\leq l< n}\;\int\limits_{\cX^2}
\left(\lam{\tW^\rT}^{l-1}\right) (x)\big[\vphi (x)p(y|x)\\
\qquad\qquad\qquad\qquad\times
\log p(y|x)\big]\left(\tW^{n-1-l}
\vphi\right) (y)\nu^2(\rd x\times\rd y)\ena\ee
and
\be\label{eq:WEkMCWPy}\beal H^\rw_{\phi_n}(f_n)=-
\int\limits_{\cX}\big[\pi (x)\log\pi (x)\big]\left(\tW^{n-1}
\vphi\right) (x)\nu (\rd x)\\
\quad -\sum\limits_{k\leq l< n}\;\int\limits_{\cX^2}
\left(\pi{\tW^\rT}^{l-1}\right) (x)\big[\vphi (x)p(y|x)\\
\qquad\qquad\qquad\qquad\times
\log p(y|x)\big]\left(\tW^{n-1-l}
\vphi\right)(y)\nu^2(\rd x\times\rd y).\ena\ee

% The contribution to \eqref{eq:WEkMC} and \eqref{eq:WEkMCe} from the
% initial terms $\log\,\pi (\BX_0^{k-1})$ and $\log\,\lam (\BX_0^{k-1})$ is negligible, as
% well as from terms containing $\log\,p(X_l|\BX_{l-k}^{l-1})$ with $l=k$ and $l=n-1$. Thus,
% our study is reduced to the analysis of  two similar expressions
% \be\label{Berhwphi1}\beal\sum\limits_{k< l< n}\;
% {\int\limits_{\cX^{n}}}\lam (\bx_0^{k-1})\vphi (\bx_0^{k-1})\big[\log\,p (x_l|\bx_{l-k}^{l-1})\big]
% \prod\limits_{i=k}^n\big[\vphi (\bx_{i-k}^{i-1})p(x_j|\bx^{i-1}_{i-k})\big]
% \nu^n(\rd\bx_0^{n-1}), \\
% \hbox{and }\;\sum\limits_{k< l< n}\;
% {\int\limits_{\cX^{n}}}\pi (\bx_0^{k-1})\vphi (\bx_0^{k-1})\big[\log\,p (x_l|\bx_{l-k}^{l-1})\big]
% \prod\limits_{i=k}^n\big[\vphi (\bx_{i-k}^{i-1})p(x_j|\bx^{i-1}_{i-k})\big]
% \nu^n(\rd\bx_0^{n-1}).
% \ena\ee \def\uvth{\underline\vartheta}

% The structure of the summand corresponding to a given $l$ is tied with the above operators $\tW$ and  $\tW^\rT$.
% Viz.,
% \be\label{eq:terml}\beal
% {\int\limits_{\cX^{n}}}\pi (\bx_0^{k-1})\vphi (\bx_0^{k-1})\big[\log\,p (x_l|\bx_{l-k}^{l-1})\big]
% \prod\limits_{i=k}^n\big[\vphi (\bx_{i-k}^{i-1})p(x_i|\bx^{i-1}_{i-k})\big]
% \nu^{n}(\rd\bx_0^{n-1})\\
% =\int\limits_{\cX_{l-k}^l}\Big[\left(\tW^{l-1}\Psi_0)(\bx_{l-k}^{l-1}\right)\Big]
% \big[p (x_l|\bx_{l-k}^{l-1})\log\,p (x_l|\bx_{l-k}^{l-1})\big]\left[\left({\tW^\rT}^{n-l-1}\u1\right)
% (\bx_{l-k+1}^l)\right] % \nu (\rd\bx_{l-k}^l)\na\ee
% where $\u1 (\bx_1^k)=1$, $\bx_1^k\in\cX^k$. (Function $\u1\in\rL_2(\cX^k,\nu^k)$ since
% measure $\nu$ has been assumed finite.) \vskip .5 truecm

At this point we use the {\it Krein--Rutman theorem} for linear operators preserving the cone of positive functions,
which generalizes  the Perron--Frobenius theorem for non-negative matrices.
The form of the theorem below is a combination of \cite{KR}, Proposition $\beta$, P. 76,
and Proposition $\beta'$, P. 77. See also \cite{D}, Theorem 19.2.

\vskip .5 truecm

{\bf Theorem (Krein--Rutman).}  {\sl Suppose that
$\cY$ is a Polish space and $\vpi$ is a Borel measure on $\cY$. Assume
a non-negative continuous kernel $K(x,y)$ satisfies the condition: $\;\exists\;$ an
integer $k\geq 0$ such that the iterated kernel satisfies the positivity condition:
$$\beal K^{(k+1)}(x,y)=\int\limits_{\cX^k}K(x,u_1)K(u_1,u_2)\cdots K(u_{k},y)
\prod\limits_{1\leq j\leq k}\vpi (\rd u_i)\geq\theta (y)>0,\; x,y\in\cY. \ena$$
Consider mutually adjoint integral operators $\tK$ and $\tK^\rT$ in
the Hilbert space $\rL_2(\cY,\vpi)$:
\be\label{eq:KK*}\beal\tK\uv (y)=\int\limits_\cY K(y,v)\uv (u)\vpi (\rd u),\;\;\uv\tK^\rT (y)
=\int\limits_\cY\uv (u) K(u,y)\vpi (\rd u)\ena\ee
and assume operators $\tK$ and $\tK^\rT$ are compact.
The following assertions hold true. {\rm{(i)}} The\\ norm $\|\tK\|=\|\tK^\rT\|:=\kappa\in (0,\infty )$
is an eigen-value of $\tK$ and $\tK^\rT$ of multiplicity one, and the corresponding eigen-functions
$\uphi$ and $\upsi$ are strictly positive on $\cX$:
$$\tK\uphi =\kap\uphi,\;\;\upsi\tK^\rT =\kap\upsi ;\;\;\uphi ,\upsi >0.$$
{\rm{(ii)}} Operators $\tK$ and $\tK^\rT$ have
the following contraction properties. Assume that $\uphi$ and $\upsi$
are chosen so that $\langle\uphi ,\upsi\rangle =1$. There exists $\delta\in (0,1)$ such that  $\forall$
function $\uv\in\rL_2(\cY,\vpi )$ with $\|\uv\|^2=\langle\uv,\uv\rangle =1$, functions $\tK^n\uv$ and
$\uv{\tK^\rT}^n$ have the following asymptotics:
\be\label{eq:vKRn}\frac{\uv\,\tK^n}{\kappa^n}=\big\langle\uv,\uphi\big\rangle
\upsi +\urQ_n,\;\;\frac{\uv\,{\tK^\rT}^n}{\kappa^n}=\big\langle\uv,\upsi\big\rangle
\uphi +\urR_n.\ee
Here $\langle\,\cdot\,,\,\cdot\,\rangle$ stands for the scalar product in $\rL_2(\cY,\vpi)$ and
the norma of vectors $\urQ_n$, $\urR_n$ are exponentially decreasing:
$$\|\rQ_n\|,\;\|\rR_n\|\leq (1-\delta )^n.$$}
\vskip .5 truecm

We are going to apply the Krein--Rutman (KR) theorem in our situation. By using the notation
$\langle\;,\;\rangle$
and $\|\;\|$ for the scalar product and the norm in $\rL_2(\cX,\nu )$, we can re-write
Eqns \eqref{eq:WEkMCWOy} and \eqref{eq:WEkMCWPy}:
\be\label{eq:WEkMCWY}\beal H^\rw_{\phi_n}(f_n,\lam )=-\Big\{
\mu\langle\upsi,\vphi\rangle\langle\uphi ,\lam\log\lam\rangle +(n-2)\langle\upsi,\vphi\rangle\,
\langle\uphi ,\lam\rangle\\
\quad\times\int\limits_{\cX^2} \uphi (y')\upsi (y)\big[\vphi (y)p(y'|y)
\log p(y'|y)\big]\nu^2(\rd y\times\rd y')\Big\}\mu^{n-2}+O((1-\delta )^n),\\
H^\rw_{\phi_n}(f_n)=-\Big\{
\mu\langle\upsi,\vphi\rangle\langle\uphi ,\pi\log\pi\rangle +(n-2)\langle\upsi,\vphi\rangle\,
\langle\uphi ,\pi\rangle\\
\quad\times\int\limits_{\cX^2} \uphi (y')\upsi (y)\big[\vphi (y)p(y'|y)
\log p(y'|y)\big]\nu^{2}(\rd y\times\rd y')\Big\}\mu^{n-2}+O((1-\delta )^n).\ena\ee
This yields that $H^\rw_{\phi_n}(f_n,\lam )\asymp\mu^n$ and $H^\rw_{\phi_n}(f_n)\asymp\mu^n$,
or, formally,
\be\label{eq:powK1}\frac{1}{n}\log\,H^\rw_{\phi_n}(f_n,\lam ),\;\frac{1}{n}\log\,H^\rw_{\phi_n}(f_n)
\to\log\,\mu.\ee
Here $\mu =\|\tW\|=\left\|\tW^\rT\right\|$ is the positive eigen-value of operators $\tW$ and $\tW^\rT$,
$\uphi$ and $\upsi$ are the positive eigen-vectors of $\tW$ and $\tW^\rT$, respectively,
as in the KR theorem. The value $\delta\in (0,1)$ represents a spectral
gap for $\tW$ and $\tW^\rT$.
% Now the statement of Theorem 5  readily follows.
\ep

We will call $\mu$ as a KR eigen-value of operator $\tW$.

\br\label{Remark 4.2.} {\rm The expressions in the curled brackets in
\eqref{eq:WEkMCWY}  do not play a role in determining the prime rate $\rB_0$. However,
they when we discuss the secondary rate $\rB_1$. Cf. Eqns
\eqref{eq:GMRepre}, \eqref{eq:GMRepres} below.}
\er

\br\label{Remark 4.3.} {\rm An assertion similar  to Theorem 6 can be proven for a general initial
distribution $\blam$ (not necessarily absolutely continuous with respect to $\nu$).}
\er

\br\label{Remark 4.4.} {\rm The Markovian assumption adopted in Theorem 6 can be relaxed
without a problem to the case of a Markov chain of order $k$.
Further steps require an extension of this techniques. See Remark \ref{Remark 4.5.} below.}
\er

The relations \eqref{eq:vKRn} in the KR theorem helps with identifying
not only the value $\rB_0$ but also
$\rB_1$ arising from a generalisation of \eqref{eq:M1} for MCs $\BX_0^\infty$ of order $k$.
More precisely, with the help of \eqref{eq:WEkMCWY} we can establish

\bt\label{Theorem 7.}
Under assumptions of Theorem \ref{Theorem 6.},
\be\label{eq:rhomm}\beal\lim\limits_{n\to\infty}{\diy\frac{H^\rw_{\phi_n}(f_n)}{n\mu^n}}
=\lim\limits_{n\to\infty}{\diy\frac{H^\rw_{\phi_n}(f_n,\lam )}{n\mu^n}}
=-\diy\frac{1}{\mu^2}\langle\upsi,\vphi\rangle\,\langle\uphi ,\pi\rangle\\
\qquad\qquad\qquad\times\int\limits_{\cX^{2}} \uphi (x)\upsi (y)
\big[\vphi (x)p(y|x)
\log p(y|x)\big]\nu^2(\rd x\times\rd y).\ena\ee.
\et

It is instructive to consider a stationary and ergodic MC, with distribution $\wt\bbP$
on $\bcX$ which os constructed as follows. The conditional and equilibrium PM/DFs
for this MC, \ ${\wt p}(y|x)
=\diy\frac{{\wt\bbP}(X_k\in\rd y|X_{k-1}=x)}{\nu (\rd y)}$ \ and
${\wt\pi}(x)
=\diy\frac{{\wt\bbP}(X_k\in\rd x)}{\nu (\rd x)}$, for $x,y\in\cX$, are given by
$${\wt p}(y|x)=\frac{W(x,y)\Phi (y)}{\mu\Phi (x)},
\;{\wt\pi}(\bx_0^{k-1})=\Psi (\bx_0^{k-1})\Phi (\bx_0^{k-1}),$$
assuming the normalization $\langle\Psi ,\Phi\rangle =\int\limits_{\cX}\Psi (x)\Phi (x)=1$.
The $n$-string PM/DF ${\wt f}_n(\bx_0^{n-1})={\wt\pi}(x_0)\prod\limits_{j=1}^{n-1}{\wt p}(x_j|x_{j-1})$
generated by ${\wt\bbP}$ has the form
$${\wt p}_n(\bx_0^{n-1})=\Psi (x_0)\prod\limits_{j=k}^{n-1}p(x_j|x_{j-k-1})\Phi (x_{n-1}).$$

The asymptotic behaviour of the WE $H^\rw_{\phi_n}(f_n)$ for a multiplicative
WF $\phi_n$ is closely related to properties important in Mathematical Physics and the theory
of Dynamical systems. In this regard, we provide here the following assertion which is known
as the variational principle for the pressure, entropy and energy. In our context, for a Markov chain
$\BX_0^\infty$ under the above assumptions,  these concepts
can be introduced in a variety of forms. Viz., for the metric pressure we can write:
\be\beacl\rB_0&=\log\,\mu =\lim\limits_{n\to\infty}{\diy\frac{1}{n}}\log\,\Xi_n\\
\;&\hbox{ where }
 \Xi_n=\int\limits_{\cX^n}\pi(x_0)\prod\limits_{1\leq j<n}W(x_{j-1},x_j)
\vphi (x_{n-1})\nu^n(\bx_0^{n-1})\ena\ee
and introduce a PM/DF $\ovp_n$:
\be\label{eq:barpin}\beac\ovp_n(\bx_0^{n-1})={\diy\frac{1}{\Xi_n}}\pi(x_0)\prod\limits_{1\leq j<n}
W(x_{j-1},x_j)\vphi (x_{n-1})\nu^n(\bx_0^{n-1}),\ena\ee
with $\int\limits_{\cX^n}\ovp_n(\bx_0^{n-1})\nu^n(\rd\bx_0^{n-1})=1$.

Note that
$$\frac{{\wt p}_n(\bx_0^{n-1})}{{\ovp}_n(\bx_0^{n-1})}=\frac{\Xi_n\Psi (x_0)}{\mu^{n-1}
\Phi (x_{n-1})}$$
and therefore
\be\label{eq:KLconve}\lim\limits_{n\to\infty}\frac{1}{n}\int\limits_{\cX^n}\log\diy\frac{{\wt p}_{n}(\bx_0^{n-1})}{
{\ovp}_{n}(\bx_0^{n-1})}\,{\wt p}_{n}(\bx_0^{n-1})\nu^{n}(\rd\bx_0^{n-1})
=\lim\limits_{n\to\infty}\frac{\log\,\Xi_n}{n}-\log\,\mu= 0.\ee

\bt\label{Theorem 8.}
Assume the conditions of Theorem {\rm 6} for the Markov chain $\BX_0^\infty$
with distribution $\bbP$. Let $\bbQ$
be a probability distribution on $\bcX_0^\infty$, with $\bbQ (\BX_0^{n-1}\in\rd\bx_0^{n-1})\prec
\nu^n(\rd\bx_0^{n-1})$ and $q^{(n)}(\bx_0^{n-1})=\diy\frac{\bbQ (\BX_0^{n-1}\in\rd\bx_0^{n-1})}{
\nu^n(\rd\bx_0^{n-1})}$, for which there exist finite rates
of the SE and the log of the kernel $W$:
\be \beac\thh (\bbQ)=\lim\limits_{n\to\infty}{\diy\frac{-1}{n}}\bbE_\bbQ\log\,q_n(\BX_0^{n-1}),\;\;
\tL(\vphi,\bbQ )=\lim\limits_{n\to\infty}{\diy\frac{1}{n}}\sum\limits_{j=k}^{n-1}\bbE_\bbQ\log\,W (X_{j-1},
X_{j}).\ena\ee
Then the quantity $\rB_0=\log\,\mu$ calculated for $\bbP$
satisfies the inequality
\be\label{eq:varprncpl} \thh (\bbQ)+\tL (\vphi ,\bbQ)\leq \rB_0.
\ee
For $\bbQ =\wt\bbP$, we have equality. Furthermore, suppose that for a stationary and ergodic
$\bbQ$ we have equality in \eqref{eq:varprncpl}. Then $\bbQ ={\wt\bbP}$.
\et

\bp The core of the argument used in the proof below is well-known in
the literature in Mathematical Physics and the theory
of Dynamical systems. We write
\be\label{eq:varineq}\beacl 0&\leq \int\limits_{\cX^n}\log\diy\frac{q_n(\bx_0^{n-1})}{{\ovp}_n(\bx_0^{n-1})}
\,q_n(\bx_0^{n-1})\nu^n(\rd\bx_0^{n-1})\;\hbox{(by Gibbs' inequality)}\\
\;&=\bbE_\bbQ\log q_n(\BX_0^{n-1})-\bbE_\bbQ\log\prod\limits_{j=k}^{n-1}W(X_{j-1},X_{j})
+\log\,\Xi_n.\ena\ee
Dividing by $n$ and passing to the limit yields \eqref{eq:varprncpl}.

Now, for $\bbQ =\wt\bbP$, we use \eqref{eq:KLconve}; this yields equality in \eqref{eq:varprncpl}.

Finally, let $\bbQ$ be a stationary process for which $\thh (\bbQ)+\tL (\psi ,\bbQ)= \rB_0$. It suffices
to check that $\forall$ given positive integer $m$, we have
$\bbE_\bbQ\tg (\BX)=\bbE_{\bbP^*}\tg (\BX )$ for any measurable and
bounded function $\tg$ depending on $\bx_0^{m-1}$. From \eqref{eq:varineq} and \eqref{eq:KLconve}
we deduce that
$$\beac\lim\limits_{n\to\infty}{\diy\frac{1}{n}}\bbE_\bbQ\log\diy\frac{q_n(\BX_0^{n-1})}{{\ovp}_n(\BX_0^{n-1})}=0
\;\hbox{ and hence }\;\lim\limits_{n\to\infty}{\diy\frac{1}{n}}\bbE_\bbQ\log
\diy\frac{q_n(\BX_0^{n-1})}{p^*_n(\BX_0^{n-1})}=0.
\ena$$
Then for $n$ large enough the ratio $f^*_{0,n-1}(\bx_0^{n-1}):=\diy\frac{q_n(\bx_0^{n-1})}{p^*_n(\bx_0^{n-1})}<\infty$
whenever $p^*_n(\bx_0^{n-1})>0$. So, $f^*_{0,n-1}$ yields the Radon--Nikodym derivative.
Moreover, setting $f^*_{m,n-1}(\bx_m^{n-1})=\bbE_{\bbP^*}\left[f_{0,n-1}(\BX_0^{n-1})|\BX_m^{n-1}
=\bx_m^{n-1}\right]$, we have that
$$\lim\limits_{n\to\infty}\bbE_{\bbP^*}\left|f^*_{0,n-1}(\BX_0^{n-1})-f^*_{m,n-1}(\BX_m^{n-1})\right|=0.$$
Then writing:
$$\beal\bbE_\bbQ\tg(\BX_0^{m-1})-\bbE_{\bbP^*}\tg (\BX_0^{m-1})=\bbE_\bbQ\left[\tg(\BX_0^{m-1})
-\bbE_{\bbP^*}\tg (\BX_0^{m-1})\right]\\
\qquad\qquad =\bbE_{\bbP^*}\left[f^*_{0,n-1}(\BX_0^{n-1})\tg(\BX_0^{m-1})
-f^*_{m,n-1}(\BX_{m+1}^{n-1})\tg(\BX_0^{m-1})\right]\\
\qquad\qquad\qquad\qquad =\bbE_{\bbP^*}\left\{\big[f^*_{0,n-1}(\BX_0^{n-1})
-f^*_{m+1,n-1}(\BX_{m+1}^{n-1})\big]\tg(\BX_0^{m-1})\right\} \ena$$
yields the desired result.
\ep

\bex\label{Example 4.1.} {\rm As an example where the above assumptions are fulfilled, consider the case
where $\cX=\bbZ_+=\{0,1,\ldots\}$, and
$\nu$ is the counting measure ($\nu (i)=1$, $i\in\bbZ_+$). The proposed transition PMF is
$$p(y|x)=\big[1-e^{-(x+1)}\big]e^{-(x+1)y},\;x,y\in\bbZ_+,$$
with the stationary PMF
$$\beal\pi (x)=\Xi^{-1}e^{-x}\big[1-e^{-(x+1)}\big],\;\;x\in\bbZ_+,\;\hbox{ where }\;
\Xi =\sum\limits_{u\in\bbZ_+}e^{-u}\big[1-e^{-(u+1)}\big].\ena$$
Conditions \eqref{eq:condHS}, \eqref{eq:condBd0}  and \eqref{eq:condL2} will be fulfilled when
we choose $\vphi\in\ell_2(\bbZ_+)$.

In a continuous setting: let $\cX=\bbR_+$, with $\nu$ being a Lebesgue measure. The transition
PDF is given by
$$p(y|x)=(x+1)e^{-(x+1)y},\;\;x,y\in\bbR_+,$$
with the stationary PDF
$$\beal\pi (x)=\Xi^{-1}{\diy\frac{e^{-x}}{x+1}},\;\;x\in\bbR_+,\;\hbox{ where }\;
\Xi =\int\limits_{0}^\infty{\diy\frac{e^{-u}\rd u}{u+1}}.\ena$$
Here conditions \eqref{eq:condHS}, \eqref{eq:condBd0}  and \eqref{eq:condL2} will be fulfilled when
we choose $\vphi\in\rL_2(\bbR_+,\nu)$.}
\eex

\br\label{Remark 4.5.} {\rm In order
to move beyond Markovian assumptions upon process $\BX=\{X_i:\,i\in\bbZ\}$, one has to introduce
conditions controlling
conditional PM/DF
$$p(y|\bx_{-\infty}^0)=\diy\frac{\bbP (X_1\in\rd y|\BX_{-\infty}^0=\bx_{-\infty}^0)}{\nu (\rd y)}.$$
At present, a sufficiently complete theory exists for the case of a compact space $\cX$, based on the theory
of Gibbs measures.
A standard reference here is \cite{Ru1}. See also \cite{Ru2}, Ch. 5.6, \cite{Ru3}, Ch. 5, \cite{Ge}, Ch. 8.3 and
the relevant bibliography therein. Extensions to non-compact cases require further work; we intend to return
to this topic in forthcoming papers. Among related papers,
Refs \cite{Su1}, \cite{Su2} may be of some interest here.}
\er

\subsection{WE rates for Gaussian processes}\label{4.C. WE rates for Gaussian processes.}

As before, it is instructive to discuss the Gaussian
case. A well-known model of a (real-valued) Markov GP $\BX_0^\infty=(X_0,X_1,\dots )$
is described via a stochastic equation
\be\label{eq:GssMrkv}
X_{n+1}=\alpha X_n+Z_{n+1},\;n\geq 0.
\ee
Cf. \cite{Ar}.
Here $\{Z_n,n\in\bbZ\}$ is a sequence of IID random variables, where $Z_n\sim\rN (0,1)$
has $\bbE Z_n=0$ and ${\rm{Var}}\,Z_n=1$. (A general case $Z_n\sim\rN (0,\sigma^2 )$
does not add a serious novelty.) The transition PDF $p(x,y)$ has the form
$p(x,y)=\diy\frac{e^{-(y-\alpha x)^2/2}}{\sqrt{2\pi}}$, $x,y\in\bbR$. The constant $\alpha$ will
be taken from the interval $(-1,1)$,
with $|\alpha |<1$. To obtain a stationary process, we take $X_0\sim\rN (0,c)$ where
$c=\diy\frac{1}{1-\alpha^2}$. This results in the (strong) solution $X_n=\sum\limits_{l\geq 0}\alpha^lZ_{n-l}$,
$n\in\bbZ$ (the series converge almost surely) and defines process $\BX$ with probability measure $\bbP$
on $\bbR^{\bbZ}$ and expectation $\bbE$. The equilibrium PDF is $\pi (x)
=\diy\frac{e^{-x^2/(2c)}}{{\sqrt{2\pi\,}}\,c}$,
$x\in\bbR$. Given $n>2$, the joint PDF $f_n(\bx_0^{n-1})=\pi (x_0)\prod\limits_{1\leq j<n}p(x_{j-1},x_j)$
for $\BX_0^{n-1}$ has the form
\be\label{eq:GssMrkvfn}\beal f_n(\bx_0^{n-1})={\diy\frac{\sqrt{1-\alpha^2\,}}{(2\pi )^{n/2}}}
\exp\bigg(-{\diy\frac{1}{2}}\bigg\{\,x_0^2 -\alpha x_0x_1\\
\quad +\sum\limits_{1\leq j\leq n-2}\diy\Big[
-\alpha (x_{j-1}+x_{j+1})x_j+ (1+\alpha^2)x_j^2\Big]
-\alpha x_{n-2}x_{n-1} +x_{n-1}^2\bigg\}\bigg)\\
\qquad =\;{\diy\frac{\sqrt{1-\alpha^2\,}}{(2\pi )^{n/2}}}\;\exp\bigg[ -{\diy\frac{x_0^2}{2}}+\alpha x_0x_1
-(1+\alpha^2){\diy\frac{x_1^2}{2}}+\alpha x_1x_2 -(1+\alpha^2){\diy\frac{x_1^2}{2}}\\
\qquad +\ldots -(1+\alpha^2){\diy\frac{x_{n-2}^2}{2}} +\alpha x_{n-2}x_{n-1} -\diy{\frac{x_{n-1}^2}{2}}\bigg]\,,
\;\bx_0^{n-1}=(x_0,\ldots ,x_{n-1})\in\bbR^n.\ena\ee

Thus, $f_n\sim\rN (\Bf0,\tC_n)$ where $\tC_n$ is the inverse of a
Jacobi $n\times n$ matrix
$$\tC_n^{-1}=\begin{pmatrix}1&-\alpha &0&\ldots &0&0\\
-\alpha &1+\alpha^2&-\alpha &\ldots &0&0\\
0&-\alpha & 1+\alpha^2&\ldots &0&0\\
\vdots&\vdots&\vdots&\ddots&\vdots&\vdots\\
0&0&0&\ldots&1+\alpha^2 &-\alpha \\
0&0&0&\ldots&-\alpha &1\end{pmatrix}\,;$$
cf. \eqref{eq:GaussPDF}. Assume that $\phi_n (\bx_0^{n-1})=\prod\limits_{0\leq j<n}\vphi (x_j)$
(a special case of \eqref{eq:phinprodk}, with $k=1$). The WE $H^\rw_{\phi_n}(f_n)=-\bbE
\phi_n(\BX_0^{n-1})\log\,f_n(\BX_0^{n-1})$ takes the form
\be\label{eq:WEGM}\beal H^\rw_{\phi_n}(f_n)={\diy\frac{1}{2}}\,\bbE\,\bigg\{\prod\limits_{0\leq j<n}
\vphi (X_j)\bigg[X_0^2-2\alpha X_0X_1+(1+\alpha^2)X_1^2\\ \qquad -2\alpha X_1X_2
+(1+\alpha^2)X_1^2
-\ldots +(1+\alpha^2)X_{n-2}^2\\
\qquad\qquad\qquad\qquad\qquad -2\alpha X_{n-2}X_{n-1} +X_{n-1}^2-2\log{\diy
\frac{\sqrt{1-\alpha^2\,}}{(2\pi )^{n/2}}}\;\bigg]\bigg\}.\ena\ee

Discarding border terms (and omitting the factor $1/2$), the bulk structure of
$H^\rw_{\phi_n}(f_n)$ is represented by the sum
$$\sum\limits_{1\leq l\leq n-2}\bbE \bigg\{
\Big[(1+\alpha^2)X_l^2-2\alpha X_lX_{l+1}+\log\,(2\pi )\Big]
\prod\limits_{0\leq j<n}\vphi (X_j)\bigg\}.$$
For a value $1<l<n-1$ away from $1$ and $n$, the corresponding summand admits
the form
\be\label{eq:GMrepre}\beal\bbE \bigg\{
\Big[(1+\alpha^2)X_l^2-2\alpha X_lX_{l+1}+\log\,(2\pi )\Big]
\prod\limits_{0\leq j<n}\vphi (X_j)\bigg\}\\
\quad ={\diy\frac{\sqrt{1-\alpha^2\,}}{(2\pi )^{n/2}}} \int\limits_{\bbR^n}e^{-x_0^2(1-\alpha^2)/2}\vphi (x_0)
\prod\limits_{1\leq j<n}e^{-(x_j-\alpha x_{j-1})^2/2}\vphi (x_j)\\
\qquad\qquad\times\Big[(1+\alpha^2)x_l^2-2\alpha x_lx_{l+1}+\log\,(2\pi )\Big]\rd x_0\ldots\rd x_{n-1}.
% \\ \quad ={\diy\frac{\sqrt{1-\alpha^2\,}}{(2\pi )^{n/2}}} \int\limits_{\bbR^n}e^{-x_0^2(1-\alpha^2)/2}\vphi (x_0).
\ena\ee
Following the spirit of the Krein--Rutman theorem we represent \eqref{eq:GMrepre} as
\be\label{eq:GMRepre}\beal \int\limits_{\bbR\times\bbR}
(\tW^{l-1}\vphi^*_0)(x)\Big[(1+\alpha^2)x^2-2\alpha xy+\log\,(2\pi )\Big]{\diy\frac{e^{-(y-\alpha x)^2/2}}{\sqrt{2\pi}}}
\left({\tW^\rT}^{n-l-2}\vphi^*_1\right)(y)\rd x\rd y\\
\quad =\mu^{n-3}\Big\{\left\langle\vphi_0^*,\upsi\right\rangle \left\langle\vphi^*_1,\uphi\right\rangle
\int\limits_{\bbR\times\bbR}\uphi (x)\upsi (y)\\
 \qquad\qquad\quad
\times \big[(1+\alpha^2)y^2-2\alpha xy+\log\,(2\pi )\big] \diy\frac{e^{-(y-\alpha x)^2/2}}{\sqrt{2\pi}}\rd x\rd y
+O((1-\delta)^{n-3})\Big\}.\ena\ee
As before, $\mu >0$ is the principal eigen-value of operator $\tW$ in $\rL_2(\bbR)$, given by
$$\tW\tf (y)=\int\limits_{\bbR}W(y,u)\tf (u)\rd u\;\hbox{ where }\;W(y,u)=\vphi (y)\diy\frac{\exp \big[-(y-\alpha u)^2/2\big]}{\sqrt{2\pi}}.$$
Next, $\uphi$ and $\uphi^*$ are the corresponding positive eigen-functions
of $\tW$ and its adjoint $\tW^\rT$, with $\tW\uphi =\mu\uphi$,
$\upsi\tW^=\mu\upsi$, $\mu =\|\tW\|=\left\|\tW^\rT\right\|$. Finally,
\be\label{eq:GMRepres}\beal\vphi_0^*(x)={\diy\frac{\sqrt{1-\alpha^2}}{\sqrt{2\pi}}}
e^{-x^2(1-\alpha^2)/2}\vphi (x),\vphi_1^*(y)={\diy\frac{1}{\sqrt{2\pi}}}\int\limits_{\bbR}e^{-(z-\alpha y)^2/2}
\vphi (z)\rd z,\;x,y\in\bbR.\ena\ee
Assuming suitable conditions on one-step WF $\vphi$, this leads to Theorems \ref{Theorem 6.} and \ref{Theorem 7.}.

\br\label{Remark 4.6.} {\rm The WE rate for a multiplicative
WF can be interpreted as a metric pressure, a concept proved to be useful in the
Dynamical system theory. The next step is to introduce a topological pressure, along
with its specific case, topological entropy. See \cite{Wa}, Ch. 9.}
\er
\def\tp{{\tt p}}

A simple example of a topological entropy and pressure in our context is as follows. Let
$\cX=\bbR$ and $\nu (\rd x)=\tp (x)\rd x$ where $\tp (x)=
\diy\frac{e^{-x^2/2}}{\sqrt{2\pi}}$. Fix a number $a>0$ and consider the set $\bcA\subset\bcX$:
$$\bcA =\{\bx =(x_i:\,i\in\bbZ ):\;|x_i-x_{i+1}|>a\;\forall\;i\in\bbZ\}.$$
Define the topological entropy $h^{\rm{top}}(\bcA ,\nu)$ by
$$h^{\rm{top}}(\bcA ,\nu)=\lim\limits_{n\to\infty}\frac{1}{n}\log\nu^n (\bcA_0^{n-1}).$$
Here
$$\bcA_0^{n-1}=\Big\{\bx_0^{n-1}\in\cX^n:\,\bx_0^{n-1}=(x_0,\ldots ,x_{n-1}):\,|x_i-x_{i-1}|>a
\;\forall\;1\leq i\leq n-1\Big\}.$$
Then $h^{\rm{top}}(\bcA ,\nu)=\log\mu$ where $\mu$ is the KR eigen-value for operator $\tW$
in $\rL_2 (\bbR)$ given by
$$(\tW\tg )(x)=\int\limits_\bbR W(x,y)\tg(y)\rd y\;\hbox{ with }\;W(x,y)=\tp (x)
{\mathbf 1}(|x-y|>a).$$
In fact, Theorem \ref{Theorem 6.} is applicable here. For the second iteration kernel
$W^{(2)}(x,y)=\int\limits_\bbR W(x,u)W(u,y)\rd u$
we have
$$W^{(2)}(x,y)=\tp (x)\int\limits_\bbR\tp (u){\mathbf 1}(|u-x|>a,|u-y|>a)
\rd u\geq c\tp (x)$$
where $c=\int{\mathbf 1}(|u|>2a)\rd u$. This implies assumption \eqref{eq:Doebl} (with $k=1$).
The Hilbert--Schmidt type condition \eqref{eq:condHS} is also fulfilled:
$$\int\limits_\bbR W(x,y)W(y,x)\rd x\rd y=\int\limits_\bbR\tp (x)\tp (y){\mathbf 1}(|x-y|>a)\rd x\rd y<1.$$

At the same time, if we set $\nu_0(\rd x)=\rd x$ then $\log\;\mu$ can be interpreted as the topological
pressure $\cP^{\rm{top}}(\bcA,\chi,\nu_0)$ for set $\bcA$, function $\chi=\ln\tp$ and reference measure
$\nu_0$:
$$\cP^{\rm{top}}(\bcA ,\chi,\nu_0)=\lim\limits_{n\to\infty}\frac{1}{n}\ln\int_{\cA_0^{n-1}}\exp\left[\;
\sum\limits_{i=0}^{n-1}\chi (x_i)\;\right]
\nu_0(\rd x_0)\cdots\nu_0(\rd x_{n-1}).$$
These connections are worth of further explorations.

%\section{Weighted rates: Markov and general ergodic sources}

\section{Rates for multiplicative Gaussian WFs}\label{Sec. 5}

In this section we focus on rates for Gaussian RPs and WFs. Recall, the SI and SE for a Gaussian PDF
$f_{m,n}=f^{\rm{No}}_{\tC_{m,n}}$ are
given by
\beq\label{eq:stGauentr} I(\bx_{m,n},f_{m,n})={\diy\frac{1}{2}}\Big\{\log\,\Big[(2\pi)^{n-m+1}
{\rm{det}}\,\tC_{m,n}\Big]
+\bx^\rT_{m,n}\tC_{m,n}^{-1}\bx_{m,n}\,\log\,\re\Big\}\eeq
and
\beq H(f_{m,n})
={\diy\frac{1}{2}}\log\,\Big[(2\pi\re)^{n-m+1}{\rm{det}}\,\tC_{m,n}\Big]
={\diy\frac{1}{2}}\Big[(n-m+1)\log\,(2\pi e)-{\rm{tr}}\;\tL_{m,n}\Big]\eeq
where $\tL_{m,n}=\log\,\tC^{-1}_{m,n}$. As before, for $m=0$, we set: $\tC_{0,n-1}=\tC_n$
and write $f_n$ for $f^{\rm{No}}_{\tC_{n}}$. (A similar
agreement will be in place for other matrices/functions emerging below.)
We can write $I(\bx_{0,n-1},f_n)={\diy\frac{1}{2}}\Big[H(f_n)-n\log\,e\Big]$.

 First, a simple example.
Suppose we take $\phi_n(\bx_{0,n-1})=e^{bn}$ where $b$ is a constant, real or complex.
(A special case of a multiplicative WF with $\vphi (x)=b$; cf. \eqref{eq:phinprodk}.)
 With $\tL_n=(L_n(i,j)$, $0\leq i,j<n)$, Eqn \eqref{eq:WEGau} becomes
\be\label{eq:WEGauebn}\beal H^\rw_{\phi_n} (f_n)={\diy\frac{e^{bn}}{2}}\Big[n\log (2\pi e)-\sum\limits_{0\leq j<n} L_n(j,j)\Big]=e^{bn}H(f_n).\ena\ee
Assume that $-{\diy\frac{1}{n}}\sum\limits_{0\leq j<n} L_n(j,j)$ converges to a value $a\in\bbR$ as $n\to\infty$.
Then ${\diy\frac{1}{n}}H(f_n)={\diy\frac{1}{n}}\Big[n\log\,(2\pi e)-{\rm{tr}}\;\tL_n\Big]\to\log(2\pi e)+a$.
Hence, we obtain ${\diy\frac{1}{n}}H^\rw_{\phi_n} (f_n)\asymp e^{bn}\big[\log(2\pi e)+a\big]$; if
$\log(2\pi e)+a\neq 0$, it impies that
$$\lim\limits_{n\to\infty}\frac{1}{n}\log\,\frac{1}{n}H^\rw_{\phi_n} (f_n)=b.$$
In general, the rate of growth/decay of $H^\rw_{\phi_n} (f_n)$ is determined by that of ${\rm{tr}}\,\tL_n$.

Next, consider an WF $\bx_{m,n}\mapsto\phi_{m,n} (\bx_{m,n})$
of the following form. Let $\tA =(A(i,j):\;i,j\in\bbZ)$ be a double-infinite real symmetric matrix
(with $A(i,j)=A(j,i)$) and assume that, $\forall$ $m<n$,
the bloc $\tA_{m,n}=(A(i,j):\,m\leq i,j\leq n)$ is such that matrix $\tC_{m,n}^{-1}-\tA_{m,n}$
is (strictly) positive definite. Then choose a real double-infinite sequence
$\bbt =(t_n,\;n\in\bbZ)$ and set
\beq\phi_{m,n} (\bx_{m,n})=\exp\,\left[\bx^\rT_{m,n}\big(\tC_{m,n}^{-1}-\tA_{m,n}\big)
\bbt_{m,n}+\frac{1}{2}
\bx^\rT_{m,n}\tA_{m,n}\bx_{m,n}\right],\eeq
where column-vectors $\bbt_{m,n}=(t_i:m\leq i\leq n)$, $\bx_{m,n}=(x_i:m\leq i\leq n)\in\bbR^{n-m+1}$.

Then the WI $\;I^\rw_{\phi_{m,n}}(\bx_{m,n},f_{m,n}):=-\phi_{m,n}(\bx_{m,n})\log\,f^{\rm{No}}_{\diy\tC_{m,n}}(\bx_{m,n})\;$ becomes
\beq\label{eq:GenGauWIm}\beal I^\rw_{\phi_{m,n}}(\bx_{m,n},f_{m,n})
={\diy\frac{1}{2}}\Big\{\log\Big[(2\pi)^{n-m+1}{\rm{det}}\,\tC_{m,n}\Big]
+\bx^\rT_{m,n}\tC_{m,n}^{-1}\bx_{m,n}\log\,\re\Big\}\\
\qquad\qquad\qquad\qquad\qquad\times \exp\left[\bx^\rT_{m,n}\Big(\tC_{m,n}^{-1}-\tA_{m,n}\Big)
\bbt_{m,n} +{\diy\frac{1}{2}}\bx^\rT_{m,n}\tA_{m,n}\bx_{m,n}\right].
\ena\eeq
To calculate the WE
$H^\rw_{\phi_{m,n}}\left(f^{\rm{No}}_{\diy\tC_{m,n}}\right)
=\int\limits_{\bbR^{n-m+1}}I^\rw_{\phi_n}(\bx_{m,n})f^{\rm{No}}_{\diy\tC_{m,n}}(\bx_{m,n})\rd\bx_{m,n}$,
we employ Gaussian integration formulas:
\beq\beal H^\rw_{\phi_{m,n}}(f_{m,n})=\int\limits_{\bbR^{n-m+1}}
\diy\frac{\log\,\Big[
(2\pi)^{n-m+1}{\rm{det}}\,\tC_{m,n}\Big]+\bx^\rT_{m,n}\tC_{m,n}^{-1}\bx_{m,n}
\log\,\re
}{2\Big[(2\pi)^{n-m+1}{\rm{det}}\,\tC_{m,n} \Big]^{1/2}}\\
\qquad\qquad\qquad\qquad\qquad\times \exp\bigg[{\diy\frac{1}{2}}\bbt^\rT_{m,n}\Big(\tC_{m,n}^{-1}-\tA_{m,n}\Big)
\bbt_{m,n}\\
\qquad\qquad\qquad -{\diy\frac{1}{2}}
\big(\bx^\rT_{m,n}-\bbt^\rT_{m,n}\big)\Big(\tC_{m,n}^{-1}-\tA_{m,n}\Big)
\big(\bx_{m,n}-\bbt_{m,n}\big)\bigg]\rd\bx_{m,n}\\
\quad =\diy\frac{\exp\left[{\diy\frac{1}{2}}\bbt^\rT_{m,n}\Big(\tC_{m,n}^{-1}-\tA_{m,n}\Big)
\bbt_{m,n}\right]}{2\Big[{\rm{det}}\Big(\tI_{m,n}-\tC_{m,n}\tA_{m,n}\Big)\Big]^{1/2}}
\Big\{H(f_{m,n})
+\diy{\rm{tr}}\Big(\tI_{m,n} -\tA_{m,n}\tC_{m,n}\Big)^{-1}\log\,\re\Big\}.\ena\eeq
\def\b0{{\mathbf 0}}

In the case $\tA=\t0$ we obtain
\beq\label{eq:expoWF}\phi_{m,n} (\bx_{m,n})=\exp\,\Big(\bx^\rT_{m,n}\tC_{m,n}^{-1}
\bbt_{m,n}\Big),\eeq
\beq \bear I^\rw_{\phi_{m,n}}(\bx_{m,n},f_{m,n})={\diy\frac{1}{2}}\Big\{H(f_{m,n})
-(n-m+1)\log\,\re
\qquad\qquad\qquad\\
+\bx^\rT_{m,n}\tC_{m,n}^{-1}\bx_{m,n}\log\,\re\Big\}\exp\Big(\bx^\rT_{m,n}\tC_{m,n}^{-1}\bbt_{m,n}\Big)\ena\eeq
and
\beq H^\rw_{\phi_{m,n}}(f_{m,n})
=H(f_{m,n})
\exp\left({\diy\frac{1}{2}}\bbt^\rT_{m,n}\tC_{m,n}^{-1}\bbt_{m,n}\right).\eeq
We arrive at a transparent conclusion. For a WF $\phi_n(\bx_{0,n-1})=\exp\,\Big(\bx^\rT_{0,n-1}\tC_n^{-1}
\bbt_{0,n-1}\Big)$ (assuming $\bbt =(t_n:n\in\bbZ)$ fixed), and given a sequence $(a(n),n\in\bbZ_+)$,
the quantity
$$\beal\hbox{$J_n(\bx_{0,n-1}):=\left[{\diy\frac{2I^\rw_{\phi_n}(\bx_{0,n-1},f_n)}{\phi_n(\bx_{0,n-1})}}
+n\log\,\re-\bx^\rT_{0,n-1}\tC_n^{-1}\bx_{0,n-1}\right]$}\\
\quad\hbox{is a constant equal to $H(f_n)$ and hence
${\diy\frac{H(f_n)}{a(n)}}\to\alpha$ \ iff \ ${\diy\frac{J_n}{a(n)}}\to\alpha$,}\ena $$
and
$$\hbox{${\diy\frac{H(f_n)}{a(n)}}\to\alpha$ \ iff \
${\diy\frac{H^\rw_{\phi_n}(f_n)}{a(n)}}\exp\left(-{\diy\frac{1}{2}}\bbt^\rT_{m,n}\tC_{m,n}^{-1}\bbt_{m,n}\right)\to\alpha$.}$$

On the other hand, for $\bbt =\b0$, the WF is simplified to
\beq\phi_{m,n} (\bx_{m,n})=\exp\,\left(\frac{1}{2}\bx^\rT_{m,n}\tA_{m,n}\bx_{m,n}\right)\eeq
whereas the WI and WE, respectively, to
\beq \bear I^\rw_{\phi_{m,n}}(\bx_{m,n},f_{m,n})={\diy\frac{1}{2}}\Big\{H(f_{m,n})
-(n-m+1)\log\,\re \qquad\qquad\qquad\\
+\bx^\rT_{m,n}\tC_{m,n}^{-1}\bx_{m,n}\log\,\re\Big\} \exp\left({\diy\frac{1}{2}}\bx^\rT_{m,n}\tA_{m,n}\bx_{m,n}\right).\ena\eeq
Furthermore,
\beq H^\rw_{\phi_{m,n}}(f_{m,n})=\diy\frac{H(f_{m,n})
+\diy{\rm{tr}}\,\left[\Big(\tI_{m,n} -\tA_{m,n}\tC_{m,n}\Big)^{-1}\right]\log\,\re}{2\Big[{\rm{det}}\,\Big(\tI_{m,n}-\tC_{m,n}
\tA_{m,n}\Big)\Big]^{1/2}}.\eeq
This implies that, for $\phi_n (\bx_{0,n-1})=\exp\,\left({\diy\frac{1}{2}}\bx^\rT_{0,n-1}\tA_n\bx_{0,n-1}\right)$,
the map $\bx_{0,n-1}\mapsto K_n(\bx_{0,n-1})$ yields a constant equal to $H(f^{\rm{No}}_{\tC_n})$. Here
$K_n$ has an expression analogous to $J_n$:
$$\bear\hbox{$K_n:=\left[{\diy\frac{2I^\rw_{\phi_n}(\bx_{0,n-1},f_n)}{\phi_n(\bx_{0,n-1})}}
+n\log\,\re-\bx^\rT_{0,n-1}\tC_n^{-1}\bx_{0,n-1}\right]$,}\qquad\\
\hbox{and hence
${\diy\frac{H(f_n)}{a(n)}}\to\alpha$ \ iff \ ${\diy\frac{K_n}{a(n)}}\to\alpha$.}\ena $$
Also,
$$\beal\hbox{${\diy\frac{H(f_n)}{a(n)}}\to\alpha$ \ iff}\\
\quad{\diy\frac{1}{a(n)}}\Bigg\{ 2H^\rw_{\phi_n}(f_n)
\Big[{\rm{det}}\,\Big(\tI_n-\tC_n
\tA_n\Big)\Big]^{1/2}
-{\rm{tr}}\,\left[\Big(\tI_n -\tA_n\tC_n\Big)^{-1}\right]\log\,\re
\Bigg\}\to\alpha .\ena$$

Similar manipulations can be performed in the general case.

\vskip 1 truecm

\noindent
{\bf Acknowledgement}
\vskip 10pt
YS thanks the Math. Department, Penn State University, for hospitality
and support. IS thanks the Math. Department, University of Denver, for support and hospitality.

\end{document}